\documentclass[12pt]{article}
\usepackage{pdflscape}
\usepackage{subcaption}
\usepackage{dsfont}
\usepackage{bbold}
\usepackage{setspace}
\usepackage{vmargin}
\usepackage{booktabs,multirow}
\usepackage[utf8]{inputenc}
\usepackage{csquotes}
\usepackage[round,sort]{natbib}
\usepackage{blindtext}
\usepackage[english]{babel}
\usepackage{fancyhdr}
\usepackage{tikz}
\usetikzlibrary{calc}
\usepackage{float}
\usepackage{svg}
\usepackage{comment}
\usepackage{amsmath}
\usepackage{stackengine}
\usepackage{amsfonts}
\usepackage{amssymb}
\usepackage{graphicx}
\usepackage{authblk}
\usepackage[normalem]{ulem}
\usepackage{vmargin}
\setmargins{2.0cm}       
{1.5cm}                  
{16.5cm}                 
{23.42cm}                
{10pt}                   
{0.5cm}                  
{0pt}                    
{2cm}                    

\usepackage{listings}
\usepackage{color} 
\definecolor{mygreen}{RGB}{28,172,0} 
\definecolor{mylilas}{RGB}{170,55,241}
\usepackage[colorlinks=true,urlcolor=blue,citecolor=blue,linkcolor=blue,bookmarks=true]{hyperref}

\begin{document}
\lstset{language=Matlab,%
    breaklines=true,%
    morekeywords={matlab2tikz},
    keywordstyle=\color{blue},%
    morekeywords=[2]{1}, keywordstyle=[2]{\color{black}},
    identifierstyle=\color{black},%
    stringstyle=\color{mylilas},
    commentstyle=\color{mygreen},%
    showstringspaces=false,
    numbers=left,%
    numberstyle={\tiny \color{black}},
    numbersep=9pt, 
    emph=[1]{for,end,break},emphstyle=[1]\color{red}, 
}

\title{ \textbf{Who benefits the most? Direct and indirect effects of a free cesarean section policy in Benin\footnote{I am grateful to Christopher Taber and Chao Fu for their feedback and advice. I also want to thank Jesse Gregory, Naoki Aizawa and my peers for their early comments and suggestions on this project.}}}

\author{Selidji Tossou\footnote{Selidji Tossou; University of Wisconsin Madison; tossou@wisc.edu}}

\affil[]{}
\date{\today}
\maketitle

\begin{center}
\end{center}

\begin{abstract}
This paper evaluates the causal effect of the access to Benin's free cesarean section policy on females and their children. I use a large sample of Demographic and Health Surveys (DHS) for West African countries and analyze how the exemption of the cesarean section user fees for females in Benin directly impacts maternal and infant mortality, family size decisions, and labor market participation. I use a Difference in Differences approach and find that having access to the free cesarean section policy significantly reduces the number of stillbirths and infant mortality by 0.0855 (a 18.79 percentage change). Second, for the surviving children, I find that access to the free cesarean section increases the likelihood of maternal mortality by 0.00465 (a 5.21 percentage change). The policy is effective at reducing infant mortality and saving the newborn. However, it harms the mother's health which translates to lower fertility after the first birth and decreased maternal labor supply post-birth. 

\end{abstract}

\clearpage

\section{Introduction}
\begin{figure}[ht]
    \centering
    \caption{Maternal Mortality Ratio, (maternal deaths per 100,000 live births)}
    \label{mortmatworld}
    \includegraphics[width=.8\textwidth]{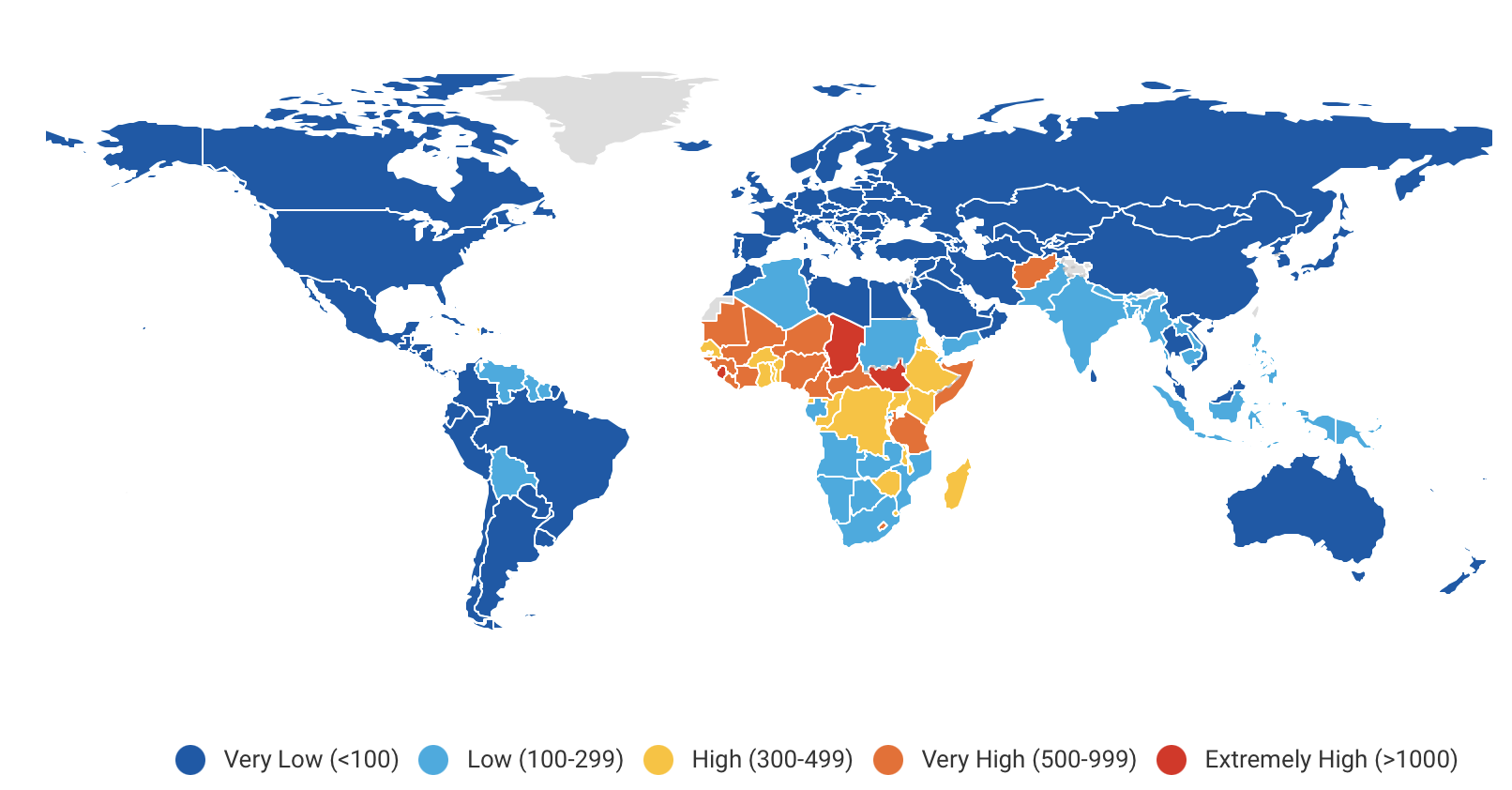}

 \uline{Source}:  \href{https://data.unicef.org/topic/maternal-health/maternal-mortality/}{World Health Organization, UNICEF, United Nations Population Fund and The World Bank, Trends in Maternal Mortality: 2000 to 2017 WHO, Geneva, 2019.} 
\end{figure}

\begin{figure}[ht]
    \centering
    \caption{Maternal mortality ratio trends by region}
    \label{mmrregion}
    \includegraphics[width=.8\textwidth]{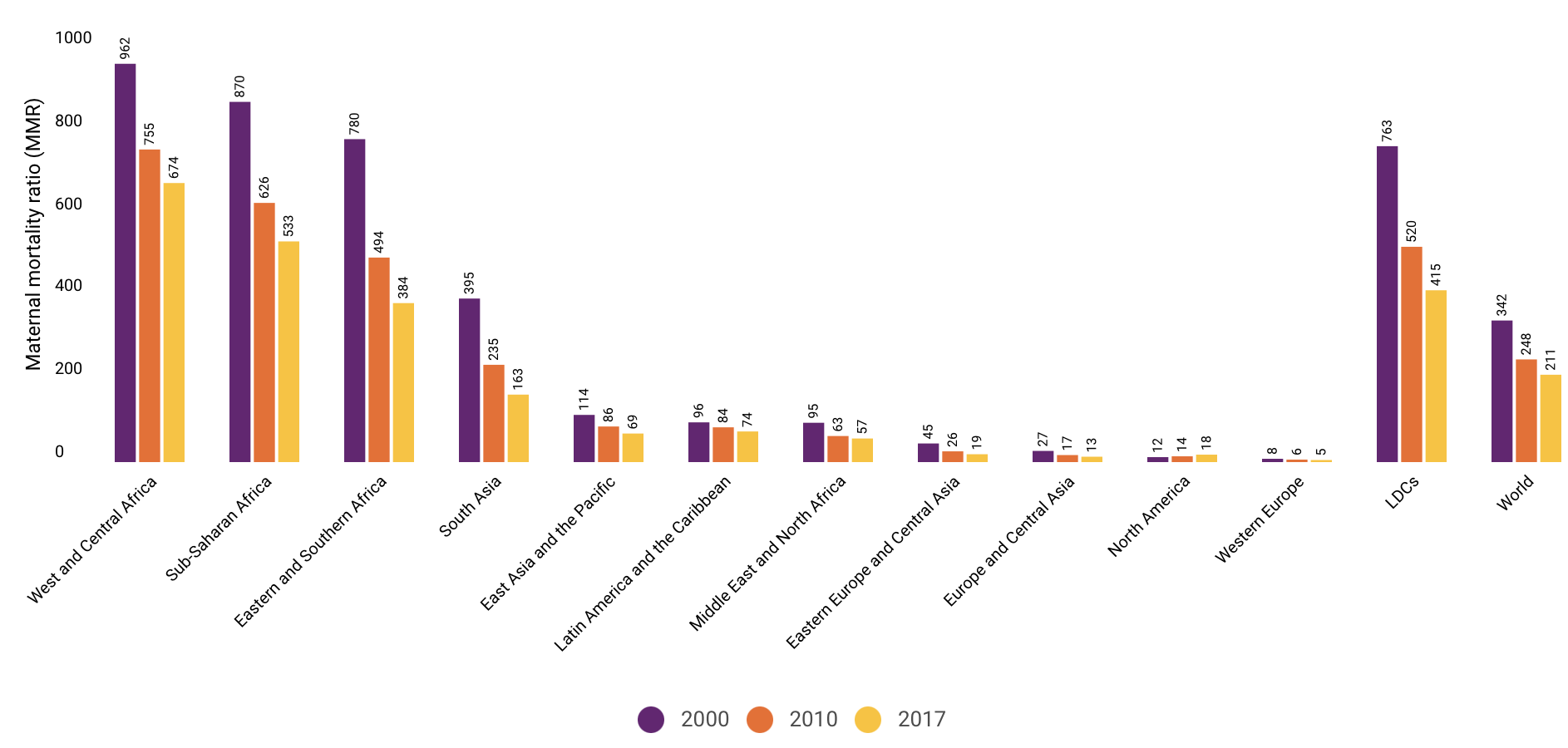}

 \uline{Source}:  \href{https://data.unicef.org/topic/maternal-health/maternal-mortality/}{World Health Organization, UNICEF, United Nations Population Fund and The World Bank, Trends in Maternal Mortality: 2000 to 2017 WHO, Geneva, 2019.} 
\end{figure}

\begin{figure}[ht]
    \centering
    \caption{Trends in pregnancy-related mortality in the United States: 1987-2017}
    \label{mort_rat_cdc}
    \includegraphics[width=.8\textwidth]{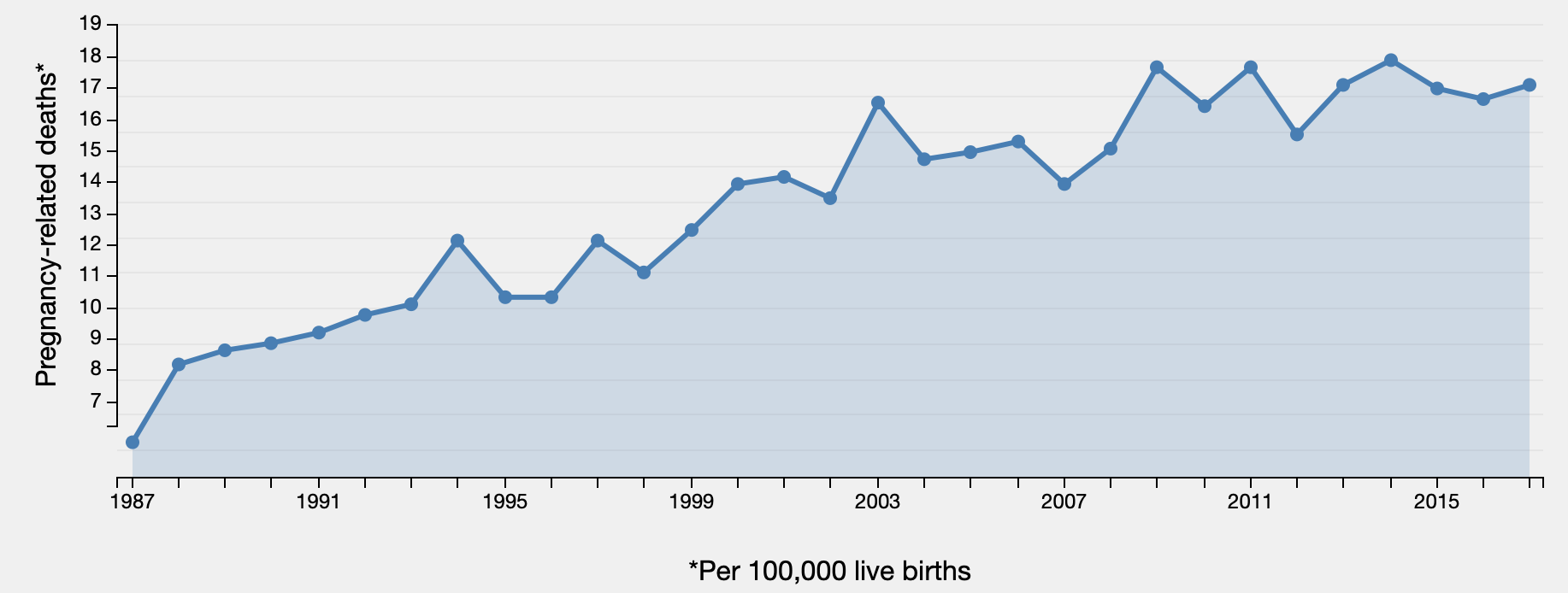}

 \uline{Source}:  \href{https://www.cdc.gov/reproductivehealth/maternal-mortality/pregnancy-mortality-surveillance-system.htm}{CDC:Pregnancy Mortality Surveillance System} 
\end{figure}

\begin{figure}[ht]
    \centering
    \caption{Causes of pregnancy-related death in the United States: 2014-2017}
    \label{mort_causes_cdc}
    \includegraphics[width=.8\textwidth]{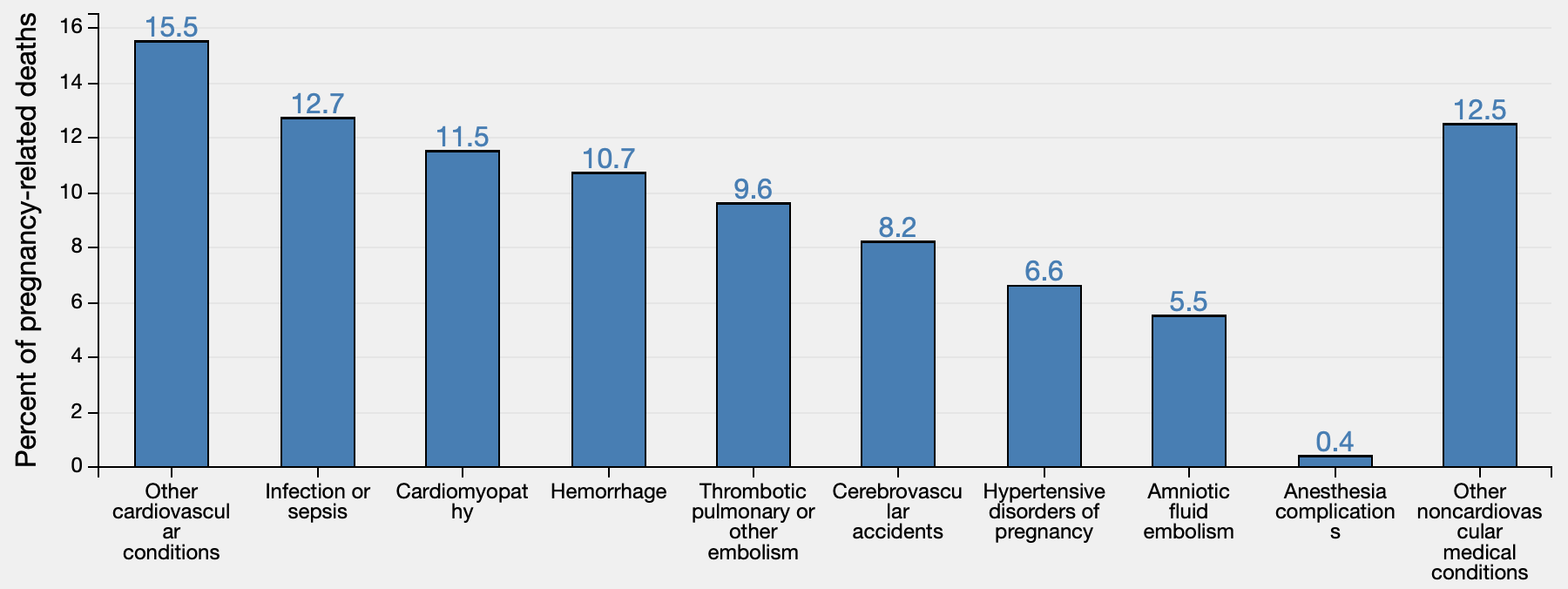}

 \uline{Source}:  \href{https://www.cdc.gov/reproductivehealth/maternal-mortality/pregnancy-mortality-surveillance-system.htm}{CDC:Pregnancy Mortality Surveillance System} 
\end{figure}

\begin{figure}[ht]
    \centering
    \caption{Benin Maternal Mortality ratio}
        \label{mortalityBenin}
    \includegraphics[width=.75\textwidth]{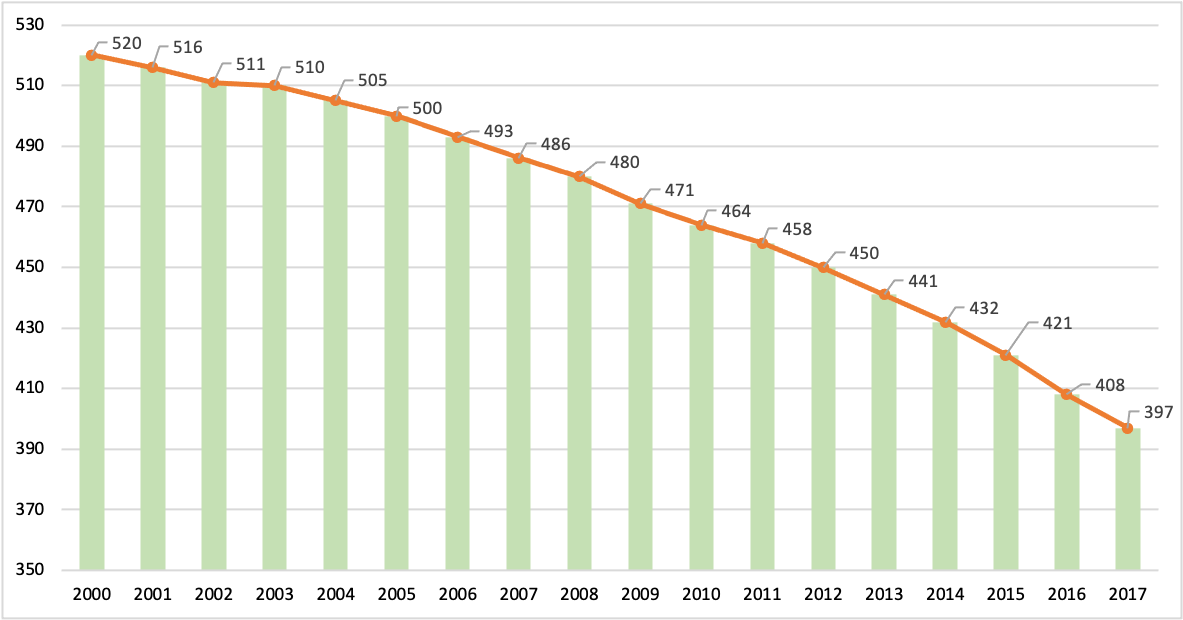}
    
    \uline{Source}: \href{https://data.worldbank.org/indicator/SH.STA.MMRT?end=2017&locations=BJ&start=2000&type=shaded&view=chart}{World Bank} 

\end{figure}

\begin{figure}[ht]
    \centering
    \caption{Under five mortality ration in Africa (modeled estimate, per 100,000 live births)}
        \label{africamortality}
    \includegraphics[width=.8\textwidth]{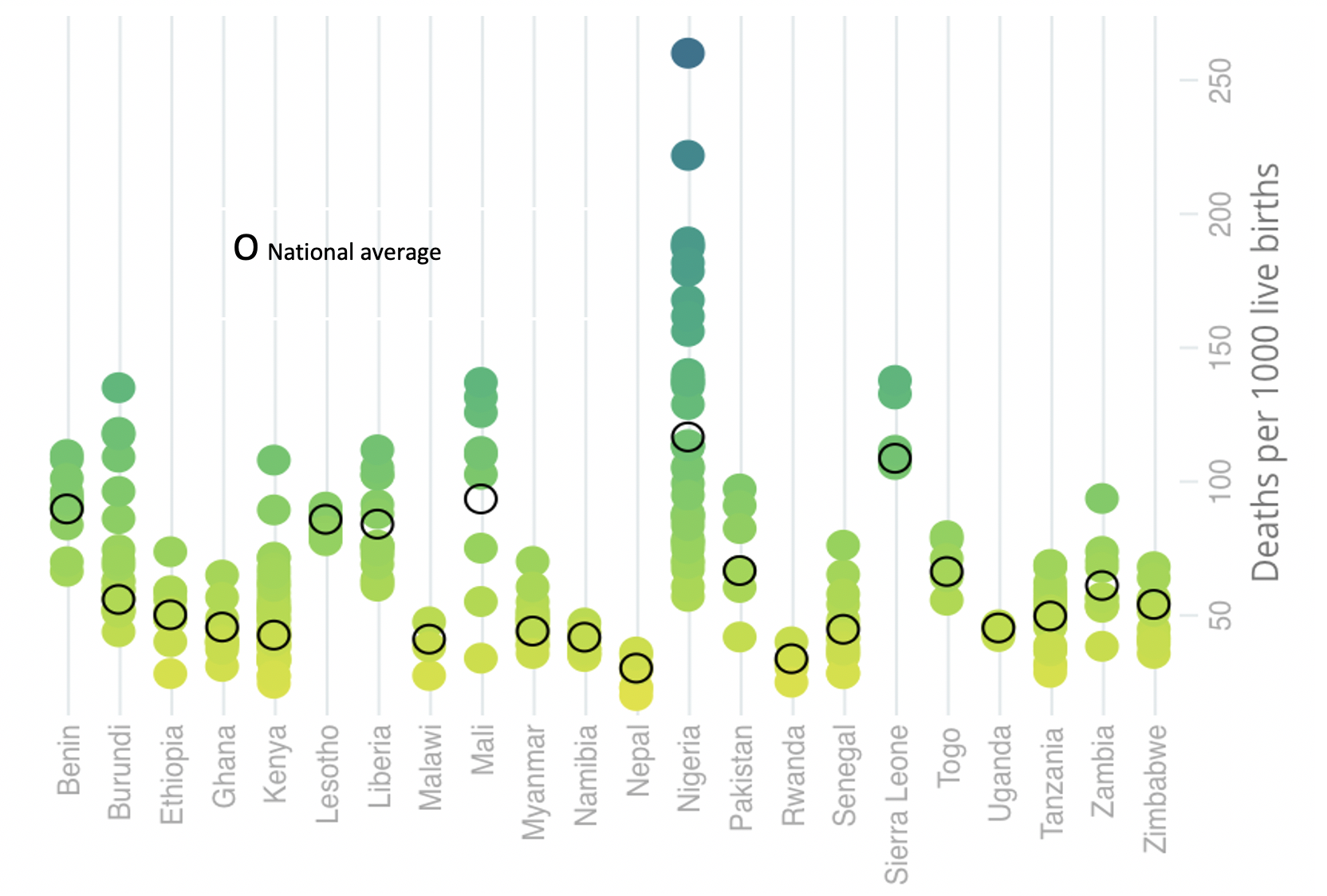}

 \uline{Source}: \href{https://data.worldbank.org/indicator/SH.STA.MMRT?end=2017&locations=BJ&start=2017&type=shaded&view=map}{World Bank} 

\end{figure}

Complications from childbirth and pregnancy can lead to maternal mortality and infant death. The World Bank defines the ratio of maternal mortality as \enquote{the number of women who die from pregnancy-related causes while pregnant or within 42 days of pregnancy termination per 100,000 live births}. In 2017, the World Health Organisation found that approximately 810 females died every day from indirect or direct preventable obstetric and gynecologic causes related to childbirth and pregnancy worldwide. Despite a drop of 38 percent in the maternal mortality ratio between 2000 and 2017, figure \ref{mortmatworld} and \ref{mmrregion} show that the issue is still severe in low-income countries where about 94 percent of the death occur \citep*{WHOmaternalmortality}. The Pregnancy Mortality Surveillance System (PMSS) was implemented in the United States to keep track of the statistics and investigate the factors and causes of deaths related to pregnancy. Figure \ref{mort_rat_cdc} shows that from 1987 to 2017, the maternal mortality ratio in the US increased from 7.2 to 17.3. The reason for the surprising increase in the maternal death ratio might include but is not limited to under-reporting and misclassification of maternal deaths, which are resolved with the improvement of identification methods and changes in the coding of causes of death\footnote{Source: \href{https://www.cdc.gov/reproductivehealth/maternal-mortality/pregnancy-mortality-surveillance-system.htm}{CDC}}. 
According to the Pregnancy Mortality Surveillance System (figure \ref{mort_causes_cdc}), these deaths are mainly reported to be caused by sepsis or infection (12.7 percent), cardiomyopathy (11.5 percent), and hemorrhage (10.7 percent). Although the global statistics are encouraging, the death of a mother or child in itself is a catastrophic situation. It can have a detrimental emotional influence on the entire family. According to the World Health Organization, there has been significant progress in resolving the maternal death issue in the last decades. In Sub-Saharan Africa, maternal mortality reduced substantially by 39 percent between 2000 and 2017. According to the World Bank, in Benin in particular, the maternal mortality ratio gradually fell from 510 deaths in 2003 to 464 deaths in 2010 and 397 deaths in 2017 per 100,000 live births (figure \ref{mortalityBenin}). In Africa, Benin is one of the deadly countries (following Nigeria and Sierra Leone) by having the third highest mortality ratio of children under the age of five (figure \ref{africamortality}). 

\begin{figure}[ht]
    \centering
    \caption{Under five mortality rate in the world: 2019}
    \label{world_under_five_mort}
    \includegraphics[width=\textwidth]{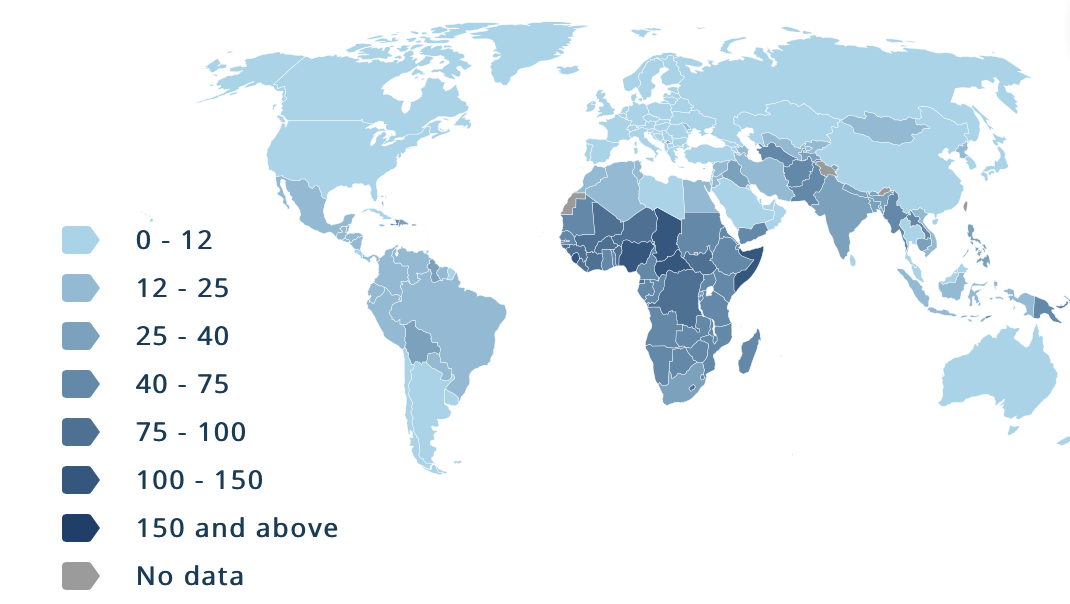}

 \uline{Source}:  \href{https://childmortality.org/data}{UN Inter-agency Group for Child Mortality Estimation} 
\end{figure}

\begin{figure}[ht]
    \centering
    \caption{Stillbirth rate in the world: 2019}
    \label{world_stillbirth}
    \includegraphics[width=\textwidth]{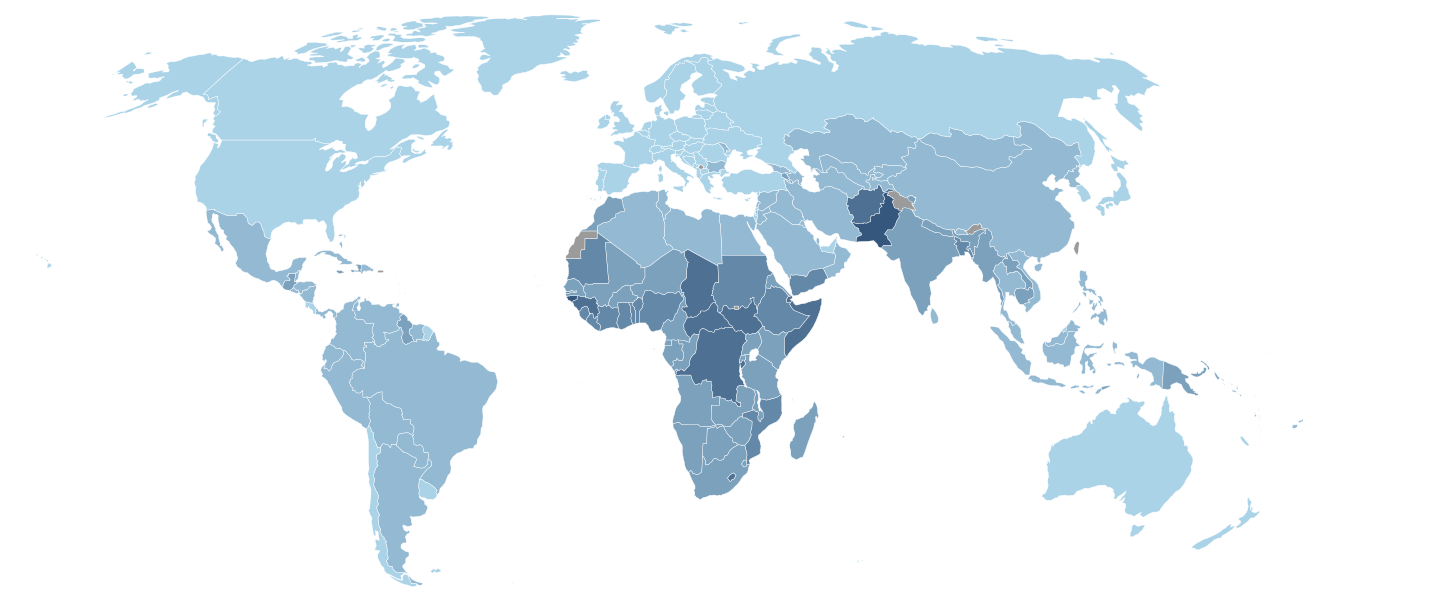}

 \uline{Source}:  \href{https://childmortality.org/data}{UN Inter-agency Group for Child Mortality Estimation} 
\end{figure}

\begin{figure}[ht]
    \centering
    \caption{Benin infant and child mortality ratios}
    \label{benin_mort}
    \includegraphics[width=.8\textwidth]{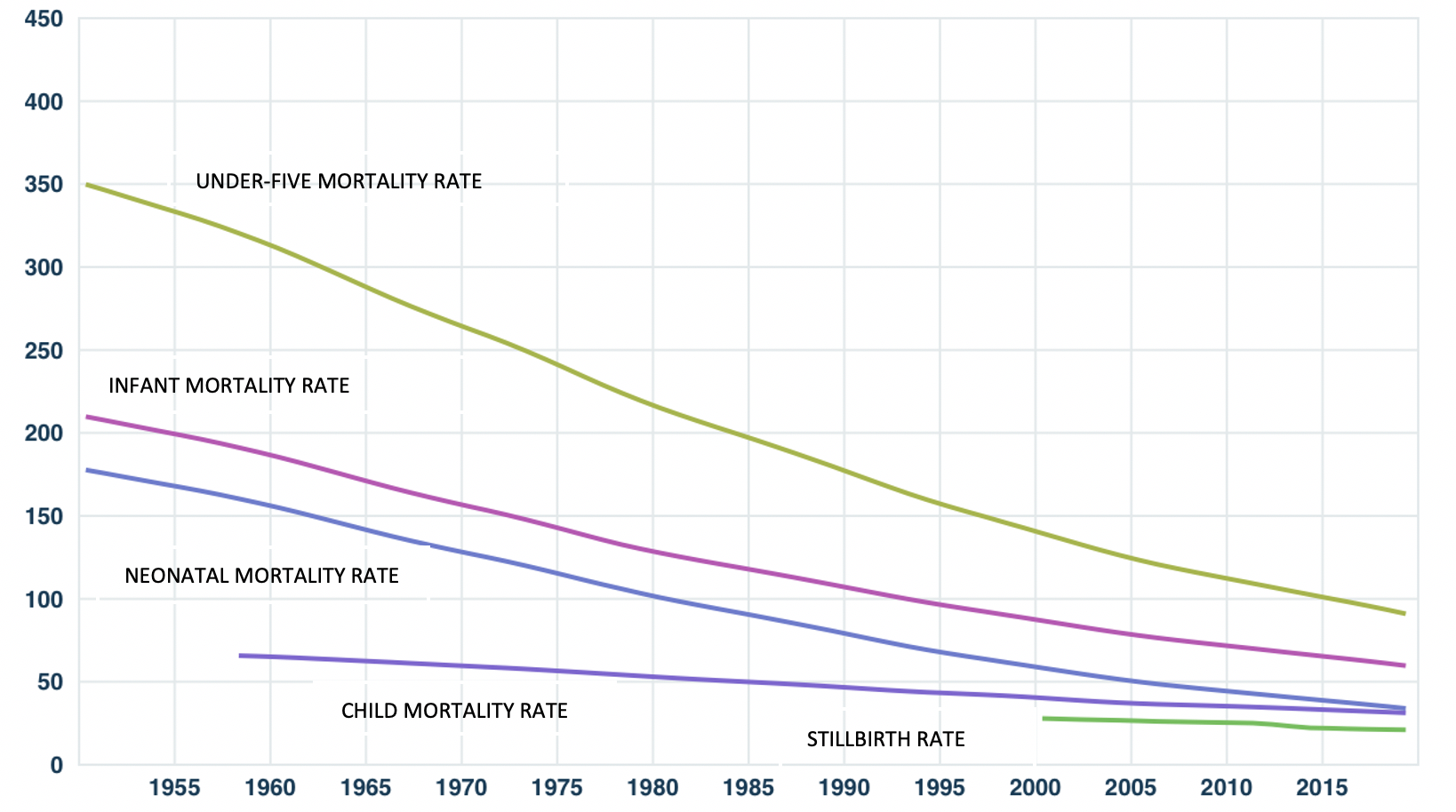}

 \uline{Source}:  \href{https://childmortality.org/data}{UN Inter-agency Group for Child Mortality Estimation} 
\end{figure}

The Centers for Disease Control and Prevention (CDC) stated many principal causes of infant death, including low birth weight and premature baby birth, injuries such as suffocation, Sudden Infant Death Syndrome, and maternal pregnancy complications. The United Nations Inter-agency Group for Child Mortality estimated that the chances of child survival are getting better worldwide. The under-five mortality rate globally decreased approximately by 59 percent from 1990 to 2019; still, more than 14,000 children die every day when they are under age 5. Figure \ref{world_under_five_mort} shows that the steepest likelihood of survival in the world is in Sub-Saharan Africa. When considering stillbirth, the United Nations Inter-agency Group for Child Mortality also estimated that worldwide, women in Southern Asia and sub-Saharan Africa bear a tremendous load of stillbirths (figure \ref{world_stillbirth}). In Benin mainly, we observe a consistent fall in the death ratio of children under age five and stillbirth (figure \ref{benin_mort}).

Since the 2000s, childbirth technologies are being developed and provided to health facilities to assist baby deliveries in sub-Saharan Africa \citep{vogel2015use}. However, some of these emergency obstetric care technologies, such as cesarean section, are costly. The average cost of a cesarean section is twofold higher than what a natural birth would cost \citep{koechlin2010comparing}. While the minimum inter-professional salary (SMIG\footnote{SMIG: Salaire Minimum Inter-Professionnel Guaranti}) was approximately US\$72 (40.000FCFA) in Benin in 2014, the cost of a cesarean section is at least US\$181 (100.000FCFA\footnote{100.000FCFA in Benin currency correspond to US\$181 using an exchange rate of US\$1 for 550 West African CFA Franc}) in Benin. Therefore, some families are still victims of maternal and infant death simply because they cannot afford such an operation. 

In Benin in 2008, President Yayi Boni announced that cesarean sections procedures in public health facilities would be subsidized through decree No 2008-730. The policy took effect in December 2008. Facilities receive approximately 100.000FCFA per operation to cover tests performed at the laboratory before the surgery, the kits, the medications, the hospitalization for up to 7 days, and transport to a better-equipped facility if necessary \citep{ravit2018removing}. Despite the advantage and concerns of cesarean section in maternal care, there is a boom in the cesarean section rate globally (In Sub-Saharan Africa;  Tanzania \citep{kabore2016determinants}, Burkina Faso \citep{litorp2015fear}. In the United States, \cite{morris2016cut} described the so-called "C-section Epidemic in America." According to the LTMII survey\footnote{LTMII: Listening to Mothers II is a survey in which a representative sample of women in the United States who gave birth was asked in 2005 questions about their labor and birth experiences.}, almost half of cesarean sections are performed on women who previously delivered by c-section, which is connected with long/short-term dangers for the newborn and the mother \citep{sandall2018short}. This situation is worse in Sub-Saharan Africa, where the probability of death after a c-section is 100 times higher \citep{sobhy2019maternal} and driven by peripartum bleeding and anesthesia complications \citep{bishop2019maternal}. Compared with countries like the United Kingdom (high-income countries) where there are only eight maternal deaths per 100,000 cesarean sections, the ratio is 100 times higher in developing countries with low and middle income. Particularly, countries in the sub-Saharan region have the highest rate of maternal deaths after cesarean sections. According to \citep{sobhy2019maternal}, data from 1990 to 2017 show that in developing countries (with low and middle income), 1 over 4 of all women who died while giving birth had done cesarean section. The question of how females are affected by Benin's free cesarean section is thus crucial to explore following these findings in the literature. 

This work is essential and relevant in many ways. Given that humans' lives are at stake, it is important to measure the direct impacts of the policy on maternal and infant mortality. In addition, in my knowledge, I am the first to tackle the question of evaluating how female labor force participation responds to the policy of free cesarean section and how females adjust their maternity decision in Benin. The article that addressed the free c-section policy in Benin looked at socioeconomic inequalities in the access to the service \citep{ravit2018removing} and found that there is no major difference in the access to cesarean sections conditional on the type of residence (urban/rural) or the education attainment of women. The question needs to be answered using a quantitative approach. It will surely contribute to policy knowledge for the policymakers in Benin in particular and Africa in general.

\section{Background}
In December 2006, the Benin Government planned to offer free care to children aged zero to five years. The Government intended to integrate pregnant women in this program for continuity of care (antenatal care, childbirth and obstetric emergencies, family planning). However, due to the high budget cost of 25 billion FCFA, they decided to start with the cesarean section fees exemption, which would cost only 2 billion FCFA. In Benin, According to UNICEF, the major causes of maternal mortality are bleeding (34 percent), infections  (10 percent), eclampsia (9 percent), and obstructed labor (4 to 8 percent). 

\begin{figure}[ht]
    \centering
        \caption{Description of Cesarean section cost in Benin}
    \includegraphics[width=\textwidth]{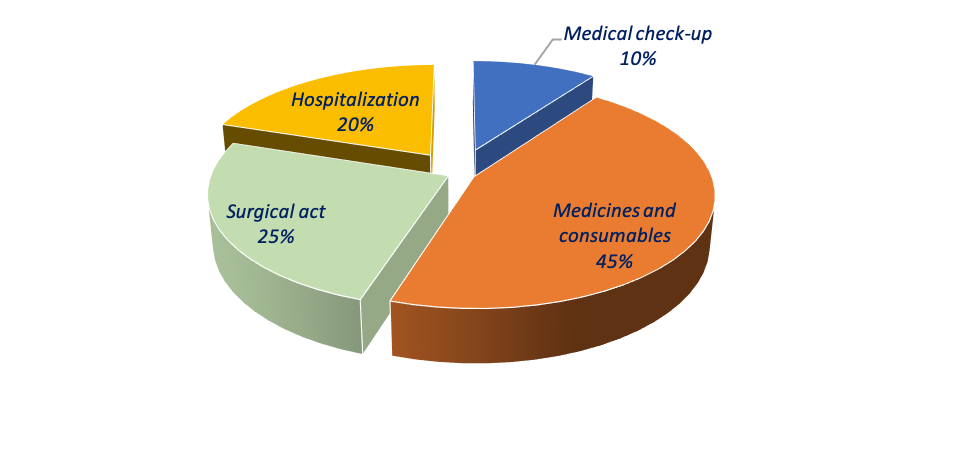}
    \label{cost}
\end{figure}

To improve financial accessibility to cesarean sections in Benin, the Government has instituted free cesarean sections through Decree No. 2008-730 of December 22, 2008. The program was announced in December 2008 and started officially in April 2009.  It began at the same time in all 43 state-approved public hospitals. A National Agency for the Management of Free Cesarean Section (ANGC) was created. The role of the ANGC is to put in place all the mechanisms necessary for the successful implementation of free cesarean sections. In particular, it is responsible for developing guidelines and reimbursing hospitals for costs incurred. 
The 43 hospitals were reimbursed monthly based on a list of cesareans, identification sheets, and invoices. 
The policy excluded private clinics and women covered by health insurance.
A lump sum of 100,000 FCFA (approximately US\$181) was retained for each cesarean section performed. The ANGC also has predefined medicines and consumables kits. Figure \ref{cost} describes the distribution of the 100,000 FCFA allowance to the health facilities. 

\begin{figure}[ht]
    \centering
        \caption{Distribution of population and hospitals in Benin}
    \includegraphics[width=.9\textwidth]{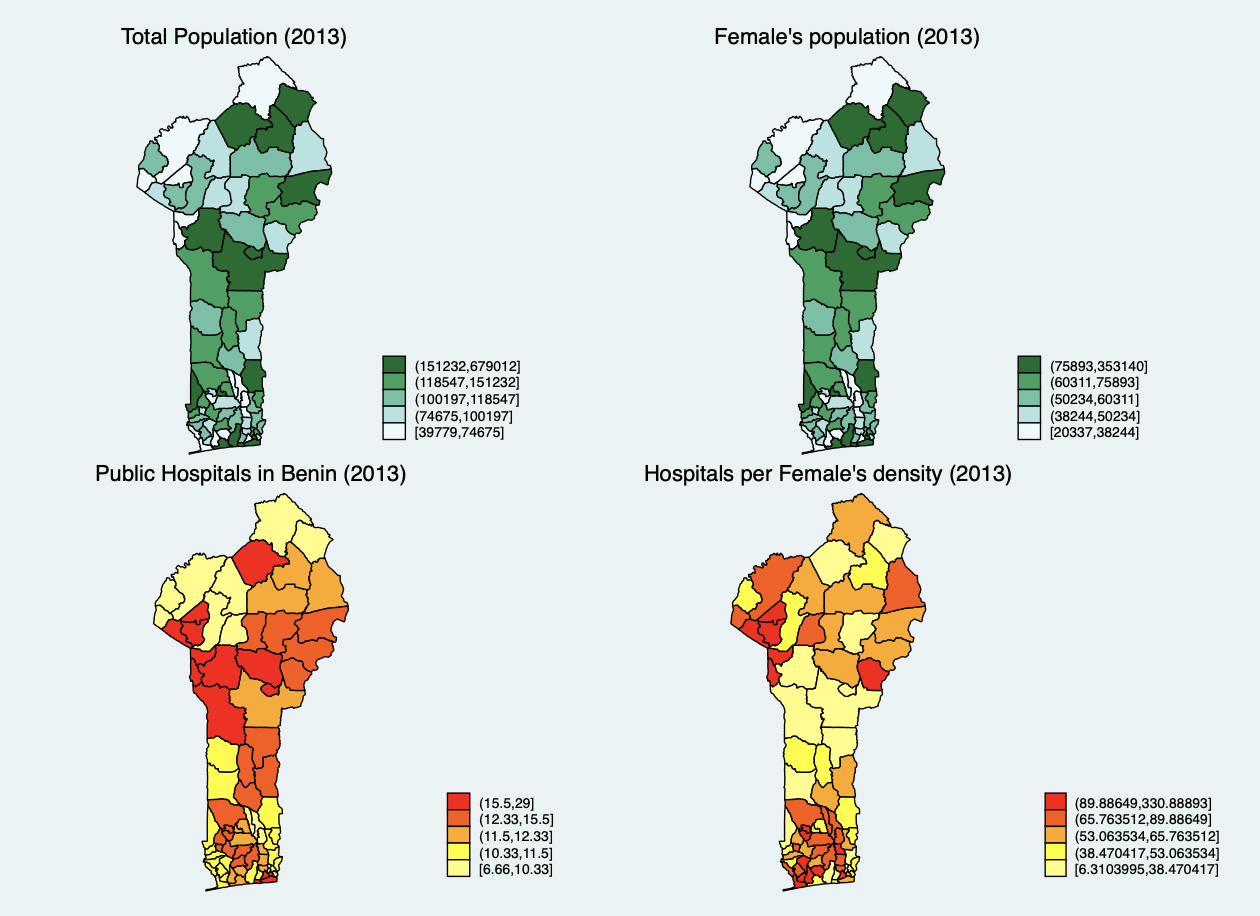}

    \small {\uline{Source}: MS/DPP/Annuaires des statistiques sanitaires and INSAE, RGPH4-2013}
    \label{Hospitals}
\end{figure}

The free cesarean section's main benefit is to remove financial barriers that limit access to quality services. Equal access to the policy was assured in many ways in Benin. First, public hospitals are present everywhere in Benin. Each department has its own facilities to provide medical services to sick people and pregnant women. The figure \ref{Hospitals} shows the distribution of public hospitals in Benin. Females would not have to pay any significant transportation costs to move to a different department or commune to access the cesarean section for free. Second, the free cesarean section policy was published through radios, televisions and newspapers as it is common for any policy that is implemented at the national level in Benin. Also, the Government recruited specific people that went through all villages and announced in each region's language to ensure that every household in Benin would have the information. Third and most importantly, pregnant women go to medical check-up for nine months. Every time they visit a hospital, the doctors ensure they are informed that the fees are exempted and paid by the Government if the delivery requires a cesarean section. 

\section{Methodology}
\subsection{Data}
I use the Demographic and Health Surveys (DHS) from the Integrated Public Use Microdata Series (IPUMS) for seven West African Countries\footnote{The seven West African Countries are Benin, Burkina Faso, Cameroon, Cote d'Ivoire, Niger, Nigeria and Togo}. The data covers the periods 2001, 2006, and 2011. The  U.S. Agency  (USAID) funds the DHS Program. They conduct large sample size household surveys that are nationally representative and allow comparisons over time. The survey targets females aged fifteen to forty-nine years and contains information on birth history, nutrition, health, postnatal care, pregnancy, and immunization. The DHS data also details domestic violence, women's empowerment, HIV/AIDS, anemia in women and anthropometry, fertility and fertility preferences, marriage and sexual activity, malaria prevention, and family planning.

Some females had access to the exemption policy of the cesarean section fees in the data, and others did not. More precisely, cesarean section is performed for free for all women in Benin if necessary. However, in Burkina Faso, Cameroon, Cote d'Ivoire, Niger, Nigeria, and Togo, the female has to pay the fees of such an operation. I separate individuals into control and treatment groups based on their access to the policy. Thus, Benin represents the treatment group directly as it comprises only females who have access to the policy, and the other countries represent the control group. Also, Benin introduced the policy in December 2008, so the data collected before 2008 (especially 2001 and 2006) are considered before the policy Implementation, and the data collected after 2008 (that is 2011) is considered after the policy Implementation.

I now discuss why I chose Burkina Faso, Cameroon, Cote d'Ivoire, Niger, Nigeria, and Togo as the control group. The DHS collects data on many other African countries with the same standard questionnaire. Thus, like Benin, I observe for more than 30 countries the same variables related to maternal health, cesarean section deliveries, and health status of many females who reported one or more births in the last three to five years before the surveys. I, therefore, have a large sample of countries to select from as a comparison group. I focus on countries with data available in the same periods as Benin to get at least three different cross-sectional datasets for more precision. Among the available countries, I focus on the most comparable to Benin regarding socio-demographic characteristics and governance. Those countries are Senegal, Mali, Niger, Ghana, Cote d'Ivoire, Burkina Faso, Nigeria, Togo, and Cameroon. I include Nigeria and Togo because of their proximity to Benin. Benin shares a border with Nigeria and Togo, and in terms of democracy and economic growth, Benin and Togo are the most comparable. I exclude Mali because the same policy of cesarean section fees exemption was introduced within the period of our study \citep{ravit2018removing}. In Senegal, the government introduced in 2005 to 2006 a policy of Maternal fees exemption with free delivery and cesarean section into some poor regions, which was later extended at regional hospital level to all areas \citep{witter2010national}. Because of that conflict, I also remove Senegal from the sample of comparison. Like many other African countries, Ghana also implemented the free maternal health program for women who are pregnant, covering any cost of health services and medicament until the delivery of the baby and after for a total of 90 days in 2008 \citep{dalinjong2017operations}. Cote d'Ivoire, Burkina Faso, and Cameroon did not have any similar policies introduced within the period of this study, and they have the DHS data available.

I code the data collected in 2001 and 2006  as pre-policy periods and data collected in 2011 as post-policy period. Table \ref{stat} gives descriptive statistics of all females in the sample. Overall, the data is well balanced and provides a good sample for comparison before and after the policy. Figure \ref{Hospitals} describes the population, female population, public hospitals, and public hospitals per female density in Benin's communes. These graphs indicate that all females have access to Public facilities. The free cesarean section policy does not discriminate over race, region, age, wealth, or individuals' observable characteristics. \cite{ravit2018removing} found that in Benin, there is no critical distinction between females living in small towns and those living in the cities, or between well-educated females and those who are not educated when considering access to cesarean sections. All cesarean sections are subject to the approval of a specialist who decide its necessity, but all females equally have equal access to Benin's fees exemption when the surgery is needed.

\subsection{Econometric Approach}
My goal is to measure the impact of the cesarean section fees exemption on infant and maternal mortality, family size decisions, and maternal labor supply in Benin. 

Let $i$ represent an individual\footnote{I have two groups of females indexed by their treatment status. Females who are not in Benin are in the Control group, and females who are in Benin are in the Treatment group. Treatment here means having access to the free cesarean section. The treatment variable $T_{c(i)t(i)}$ simply takes value $1$ for Benin after the policy implementation, which is in 2008. Every female is indexed by the letter $i=1,...,N$.}, $c(i)$ the country where individual $i$ is located, $t(i)$ the time period I observe the individual $i$ and $$T_{c(i)t(i)}= 
 \begin{cases} 
 1, & \mbox{if } c(i)= \mbox{Benin and } t(i)\geq 2008 \\ 
 0, & \mbox{  otherwise} 
 \end{cases} $$

My main specification is the Difference in Differences in which the effect of the treatment on an outcome variable $Y_i$ is evaluated through the following standard equation:
\begin{equation}
Y_i = \beta_0 + \beta_1 T_{c(i)t(i)} + \gamma_{c(i)} + \rho_{t(i)} + X_i^{'}\alpha + \mu_i
\end{equation}

where $\beta_0, \alpha, \beta_1$, are unknown parameters and $\mu_i$ is an unobserved random variable which includes all determinants of the outcome variable $Y_i$ omitted in the model. $\beta_1$ is the main coefficient of interest in the model and reveals the true effect of the treatment. By evaluating the access to the free cesarean section, the ultimate goal is to find a good estimate $\hat{\beta_1}$ of $\beta_1$ using the data available. 

\begin{figure}[ht]
    \centering
    \caption{Parallel trends assumption for the average number of children}
    \label{trendTC}
    \includegraphics[width=.7\textwidth]{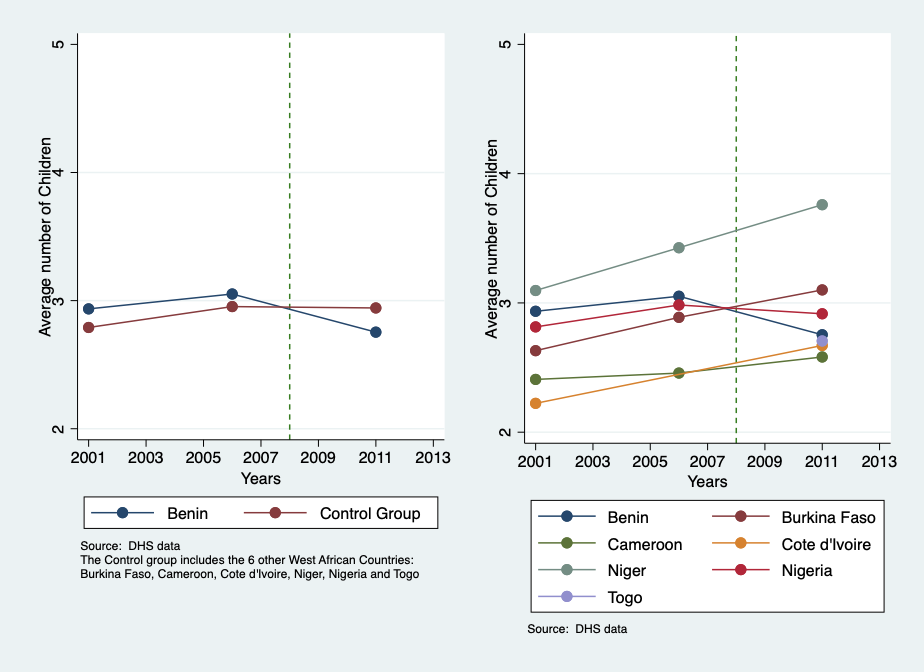}
    
    \footnotesize \uline{Note}: TC represents the total number of children a female has; the vertical dashed line represents the introduction of the policy in Benin in December 2008. There is a parallel pattern between the two groups before the policy implementation.
\end{figure}

\begin{figure}[ht]
    \centering
        \caption{Parallel trends assumption for the maternal labor supply}
    \label{trendLabor}
    \includegraphics[width=0.7\textwidth]{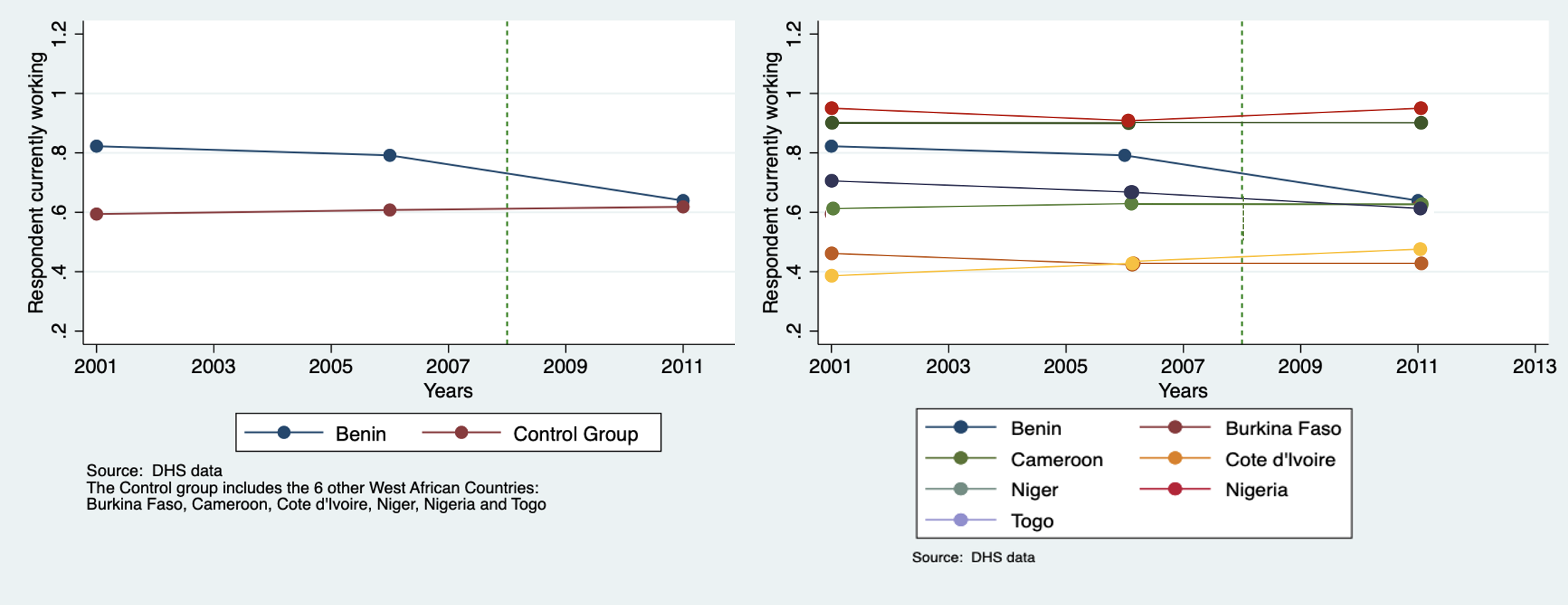}
    
    \footnotesize \uline{Note}: The outcome of interest is maternal labor supply. There exists a similar trend between the two groups before the implementation of the policy
\end{figure}

\begin{figure}[ht]
    \centering
        \caption{Parallel trends assumption for the Infant mortality}
    \label{infant}
    \includegraphics[width=0.7\textwidth]{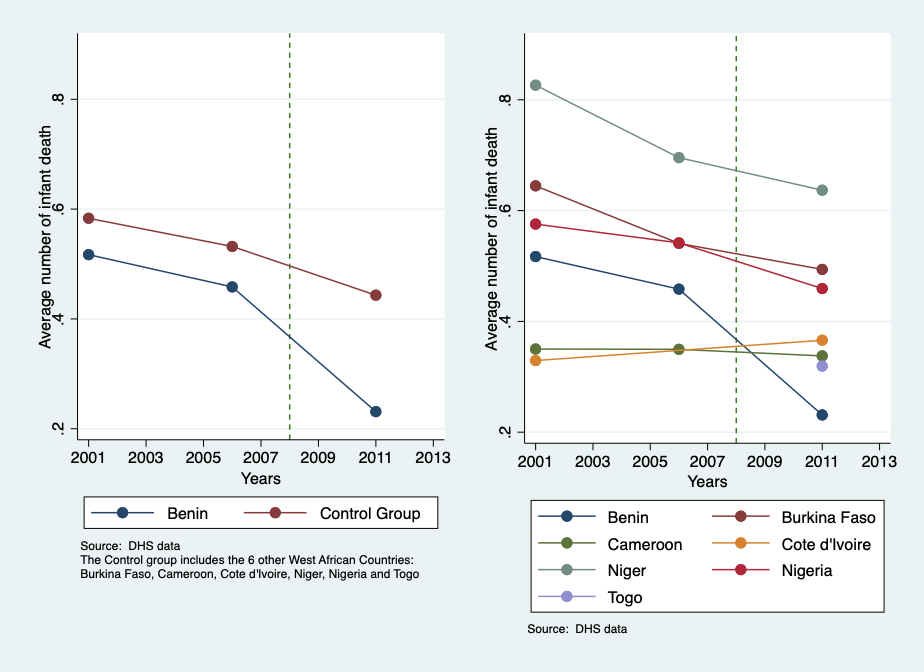}
    
    \footnotesize \uline{Note}: The outcome of interest is infant death. It also include Pregnancy lost due to stillbirths and miscarriages. There exists a similar trend between the two groups before the implementation of the policy
\end{figure}

Some key assumptions must hold to estimate a causal effect using a Difference in Differences approach. First, the intervention must be unrelated to the outcome at baseline. In Benin's case of free cesarean section, the employment status or the number of children of an individual does not determine the intervention's allocation. All females equally have access to the policy regardless on their socio-demographic or economic conditions. Second, the control and treatment groups must satisfy the parallel trends Assumption in the variable. The parallel trends assumption is the most important assumption to ensure the Difference in Differences model's internal validity and is the hardest to fulfill. This assumption requires that the gap between the comparison group and the group that received the treatment is the same over time without the treatment. 
The first column of figure \ref{trendTC} for example, shows the trends of women's fertility for the two groups before and after the intervention in 2008. There is approximately a parallel pattern between the two groups before the policy. Individuals who would not have benefited from the free cesarean section after the policy tend to have fewer children than the other group. This gap changed after the intervention. I also notice that Benin is the only country where there is a large change in the number of children after 2008 (right column of figure \ref{trendTC}). These findings suggest that the access to the free cesarean section introduced in 2008 may have affected the decision of females to have more children. 
I also verify the parallel trends assumption for maternal labor supply and infant mortality in figure \ref{trendLabor} and figure \ref{infant} respectively. Figure \ref{trendLabor} shows that there exists a similar trend between the two groups before the implementation of the policy for the maternal labor supply. Finally, figure \ref{infant} also shows a parallel direction between Benin and the control group for infant death and pregnancy loss. I notice a significant drop in the death rate for Benin after the implementation of the policy, while other countries kept approximately their trends in the control group. Third, the constitution of the control group and treatment group must be stable for repeated cross-sectional research. In this case, the descriptive statistics show a comparable group of individuals across the different years. 

The DHS data is a repeated cross-sectional data. It provides information only on females who are alive. Therefore, I cannot use that data to evaluate maternal mortality. Instead, I use Census data from the Integrated Public Use Microdata Series (IPUMS) for Benin as the treatment group and the six other West African countries are in the control group. The census data however provides the mortality status of mothers only for children who are alive. Therefore, due to selection issues, I cannot profoundly analyze the impact of the policy on female mortality. I evaluate the maternal mortality for the surviving kids and provide descriptive analyses using aggregate data on females in Benin. 

\subsubsection{Infant Mortality}

I use the following equation to measure the direct impact of the access to the free cesarean section on infant death and stillbirth in Benin.

\begin{equation}
    Number\_Death_i = \beta_0 + \beta_1 T_{c(i)t(i)} + \gamma_{c(i)} + \rho_{t(i)} + X_i^{'}\alpha + \mu_i
\end{equation}

Here, $Number\_Death_i$ includes for each female $i$ in the data the number of miscarriages, stillbirths, and infant deaths within the 3 years till the data collection. While miscarriage and stillbirth both represent a pregnancy's loss, they differ according to when the loss occurs. Miscarriage is when a baby in the womb dies before the pregnancy reaches 20 weeks; stillbirth is when a baby in the womb dies at or after the pregnancy reaches 20 weeks. Infant death here simply represents the loss of a child before the age of five. For an easier interpretation, I also measure the infant mortality as the likelihood of infant death through the equation:

\begin{equation}
    Infant\_Death_i = \beta_0 + \beta_1 T_{c(i)t(i)} + \gamma_{c(i)} + \rho_{t(i)} + X_i^{'}\alpha + \mu_i
\end{equation}
 where $Infant\_Death_i$ is a binary variable indicating if a female $i$ ever had an infant death or pregnancy loss within the 3 years till the data collection. $T_{c(i)t(i)}$ represents females in Benin after the implementation of the free cesarean section policy in 2008. $\gamma_{c(i)} $ and $ \rho_{t(i)}$ represent country fixed effect and time fixed effect respectively. $\beta_1$ is the main coefficient of interest and $X_i^{'}$ consists of the variables I control for such as the age of the female, age of partner, type of residence (urban/rural), education, wealth index, etc. 

\subsubsection{Maternal Mortality}

I use the following equation to measure the direct impact of access to the free cesarean section on maternal mortality in Benin.

\begin{equation}
Mother\_Death_i = \beta_0 + \beta_1 T_{c(i)t(i)} + \gamma_{c(i)} + \rho_{t(i)} + X_i^{'}\alpha + \mu_i
\end{equation}

Here, $Mother\_Death_i$ is a binary variable that indicates the death status of the mother in each household in the Census data. I restrict the household to the youngest kid\footnote{I restrict the data to one individual per household in order to get one binary variable for all siblings for the death status of the mother. This will avoid counting many time the same mother in the analysis.} and I use children of age five or less and interpret the result as the effect of the access to the exemption of cesarean section user fees on the likelihood of death of the mother for the surviving kid. The equation also includes a vector of country fixed effects $\gamma_{c(i)}$. Additionally, in order to control for unobserved heterogeneity over time, I add time fixed effects $\rho_{t(i)}$. I estimate the equation by ordinary least squares as a linear probability model\footnote{I estimate the likelihood of maternal mortality with the logit model and obtain the same result}. The coefficient of interest $\beta_1$ captures the relationship between the access to the free cesarean section and the likelihood of maternal mortality for the surviving child. A positive coefficient would suggest a negative effect of the access to the free cesarean section on maternal mortality, probably due to cesarean sections being both over and underused, which can contribute to adverse outcomes. Complications occurring during or following a cesarean section can be particularly severe in developing countries (with low and middle income) due to the lack of resources and trained personnel needed to manage the complications.

\subsubsection{Family Size}

I use a Difference in Differences model to measure how the policy affected family size decision through the model :
\begin{equation}
Number\_Children_i = \beta_0 + \beta_1 T_{c(i)t(i)} + \gamma_{c(i)} + \rho_{t(i)} + X_i^{'}\alpha + \mu_i
\end{equation}

where $Number\_Children_i$ represents the total number of children, which is the outcome of interest. $X_i^{'}$ represents other covariates I control for, such as the age of the female, age of partner, type of residence (urban/rural), education, wealth index, etc. Age is crucial for my identification strategy. The age of men and women affects fertility. By fertility, I mean the ability of a female to get pregnant and have children. Age is a significant factor that affects a woman's chance to fall pregnant, carry the pregnancy to the full term, and conceive a healthy baby. A woman's fertility starts to decline toward the 30s and reduce even more after the age of 35. The risks of carrying a pregnancy with complications increase with the age of women. My model, which includes female age as a covariate, evaluates whether a female has more children conditional on current age. 

Type of residence also plays an essential role in explaining fertility \citep{WhiteMJ}. In my setting, the type of residence is divided into two groups: urban area and rural area. \citet{kulu2013fertility} examined fertility variation across settlements and found that the level of fertility tends to be more in small towns compared to developed areas or capital cities. \citet{glusker2000differences} observed the same pattern in Washington.  \citet{becker1991treatise} argued that compared to rural areas, urban zones favor higher financial investment on children (through urban areas' norms, the closeness of children to attractions, extra-curriculum activities, etc.), which are finally higher than any gap in the earnings between urban and rural areas. \citet{michielin2004lowest} also argued that type of residence affects fertility mainly because, compared to rural areas, the cost of living is higher in urban areas. Another important covariate I use is education. \citet{martin1995women} studied the relationship between the education level of women and their fertility and noticed that educational attainment is negatively correlated with fertility. The author explains that education empowers women and develops their ability to make their own choices regarding reproduction, which does not increase the demand for children. The wealth index may also be an important variable for understanding family size decisions. Though the effect is ambiguous, \citet{colleran2018variation} found that the effect of wealth on fertility varies across populations, while the effect of education on fertility is negative constantly. A negative effect on fertility can indirectly consolidate the argument of the adverse effect of the cesarean sections on females' health in case they survive the surgery.

\subsubsection{Maternal Labor Supply}

In the data, females were asked whether they were currently employed or not. Thus, this binary variable takes the value one if the female is presently working or 0 if not. Therefore, what I want to evaluate is, for a female who had access to the exemption of cesarean section fees (versus the female who did not), what is the probability that she will be currently working? First, I started with a linear probability model as follows:

\begin{equation}
Maternal\_Labor\_Supply_i  = \beta_0 + \beta_1 T_{c(i)t(i)} + \gamma_{c(i)} + \rho_{t(i)} + X_i^{'}\alpha + \mu_i
\end{equation}

In this model, I simply applied the Ordinary Least Square to the Maternal Labor Supply variable ($Maternal\_Labor\_Supply_i$), a binary outcome instead of a continuous variable.
As in the family size model, I evaluated the maternal labor supply conditioning on the same set of covariates (age of the female, age of partner, type of residence (urban/rural), education, wealth index). There exist extensive literature on how each of those variables affects labor supply. These variables play an important role in my identification strategy.

Using linear probability model is advantageous because it is easy to interpret the coefficients of estimations. However, Linear Probability Models are not always perfect \citep{tossou2025essays}. Three main issues can arise. Although these problems do not essentially make my analyses wrong, the first issue is the non-normality of the error term. The second problem is heteroskedasticity of the errors, and the most important is potentially nonsensical predictions. After estimating the econometric model, the assumption that the error follows a normal distribution is most critical for performing hypothesis tests. However, the error term of a LPM follows a Distribution which is binomial. 
Second, in the linear regression model, the homoskedasticity of the error term is assumed. This is required to show that the estimators of the Ordinary Least Square are efficient (or best). In a Linear Probability Model, the error term is not homoskedastic. Lastly, the main law of probability requires that the likelihood of an event must be in the interval $[0,1]$. However, the nature of a Linear Probability Model does not guarantee the fulfillment of this requirement. I can run a logit model as a sensitivity analysis because some predicted probability from the Linear Probability Model might be out of the range of $[0,1]$. The Logit model forces the probability to be bounded between 0 and 1. However, in the context of Difference in Differences, the resulting coefficients are
hard to interpret. The estimates in the logit model are interpreted as the log odds of the outcome taking value one. 


\section{Findings}
\subsection{Infant Mortality}

\begin{figure}[ht]
    \centering
    \caption{Cesarean section use}
    \includegraphics[width=.8\textwidth]{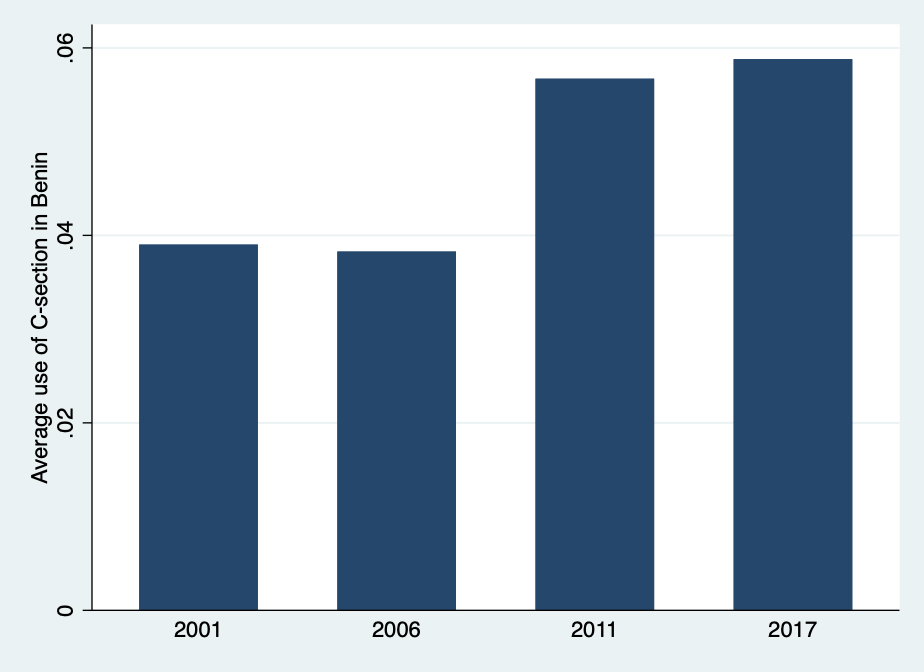}
    
 \uline{Source}: {DHS data} 
    \label{csecuse}
\end{figure}

I first investigate the infant death rate under the policy.  
Figure \ref{csecuse} shows that the exemption of cesarean section fees in Benin leads to an increase in the use of the service. What are the implications for infant mortality? I use a Difference in Differences specification to explore the possible impacts of access to the free cesarean section on infant mortality. The result in table \ref{mortinfant11} suggests that the access to the policy significantly reduced the number of infant death and stillbirth in Benin by 0.0855 (this represents a percentage change of 18.79 percent\footnote{I divide the estimate of beta by the mean of the dependent variable: $-0.0855/0.455=-0.1879$}). This means that for over 100 newborns who would have died in the absence of the policy, approximately 18 are saved because of the access to the exemption of the cesarean section user fees. This result is promising and aligns with the goals of the free cesarean section policy, which primarily intends to lower the infant mortality in Benin. To the best of my knowledge, this is the first paper that evaluates the causal effect of free cesarean section policy on Benin's infant mortality. Wealth and education attainment play a role in decreasing infant mortality. This can be explained by the fact that the free cesarean section covers only the surgery and a medicine kit for the mother. It does not cover prenatal and antepartum checkups. So a pregnant woman who has the means is more likely to identify and prevent any problems related to the pregnancy and the newborn's health. Table \ref{mortinfant21} gives an easier interpretation of the effect of the access to the policy on infant mortality. In this table, the outcome of interest is a binary variable that takes value 1 if a female has an infant death or a pregnancy loss in the three years preceding the data collection and 0 otherwise. Thus, I interpret the coefficient $\hat{\beta_1}$ as the effect of the policy on the likelihood of infant mortality. Using my most preferred specification (Column (4)), the result suggests that access to the policy decreases the likelihood of infant death or pregnancy loss of a female in Benin by 0.0962. Compared to the mean of the dependent variable, this effect represents a percentage change of 36.85 percent. To see which female category is driving this result, I decompose the effect by age group, wealth index, and education attainment. 

The first panel of table \ref{mortinfant31} presents the stratification by age group. The policy is more effective at reducing infant mortality for females older than 25. For females between 15 and 24, the probability of infant mortality is reduced by only 4.65 percent compared to a reduction of 30.85 percent for a female between 25 and 34, a reduction of 43.53 percent for females between 35 and 44, and the highest decrease of 47.25 percent for females between 45 and 49\footnote{These are the percentage change of infant mortality obtained by dividing the coefficient estimated by the mean of the dependent variable}. The same exercise is done by using the wealth index of females. The results in the second panel of table \ref{mortinfant31} suggest that the effect of the free cesarean section on infant mortality is the biggest for females who are the poorest, poorer, or have a middle wealth index\footnote{The modalities Poorest, Poorer, Middle, Richer, and Richest are pre-computed in the DHS data} (46.36 percent, 35.84 percent, and 41.72 percent respectively). For the richer and richest females, the effect of the policy is a percentage change reduction of 28.52 percent and 16.47 percent on the likelihood of infant death. The last panel of table \ref{mortinfant31} gives a stratification by education level. The highest effect of the policy on the reduction of the likelihood of infant death (-39.07 percent) is for females who have no education level. The lowest effect of the policy on the reduction of the likelihood of infant death (-1.4 percent) is for females who have higher education levels.\footnote{I test the difference between all the coefficients and they are all significant and statistically different} 

To summarize, I find that access to the free cesarean section in Benin reduces the likelihood of infant mortality on average by 36.85 percent. The policy is more effective at reducing infant death for older females, females who are below the middle wealth index, and females who have no education. 

\subsection{Maternal Mortality}

\begin{figure}[ht]
    \centering
    \caption{Number of births per 1,000 women per age in Benin in 2020}
    \includegraphics[width=.8\textwidth]{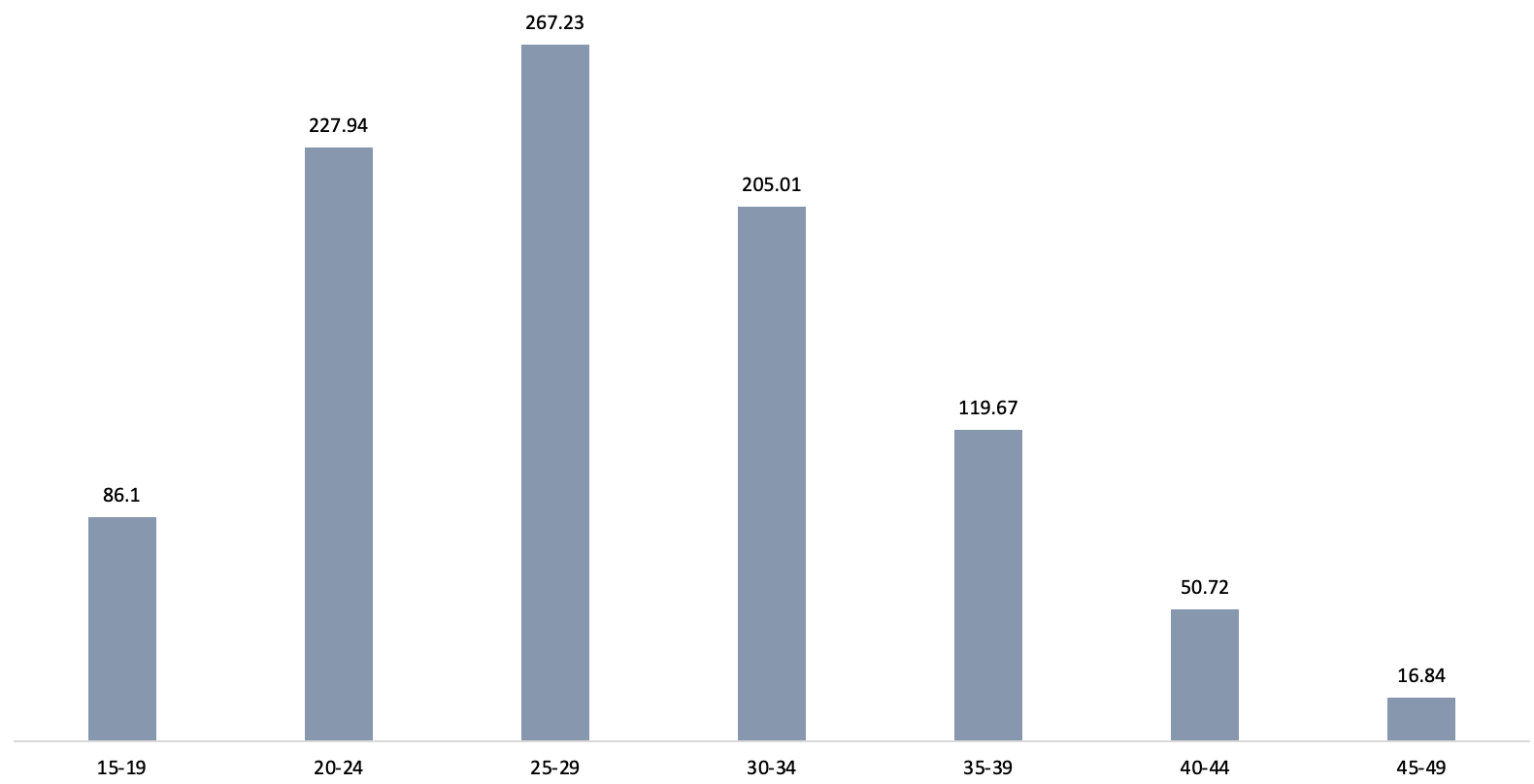}
    
 \uline{Source}: {World Bank} 
    \label{fertilitybenin}
\end{figure}

\begin{figure}[ht]
    \centering
    \caption{Age distribution of females before and after the policy}
    \includegraphics[width=.8\textwidth]{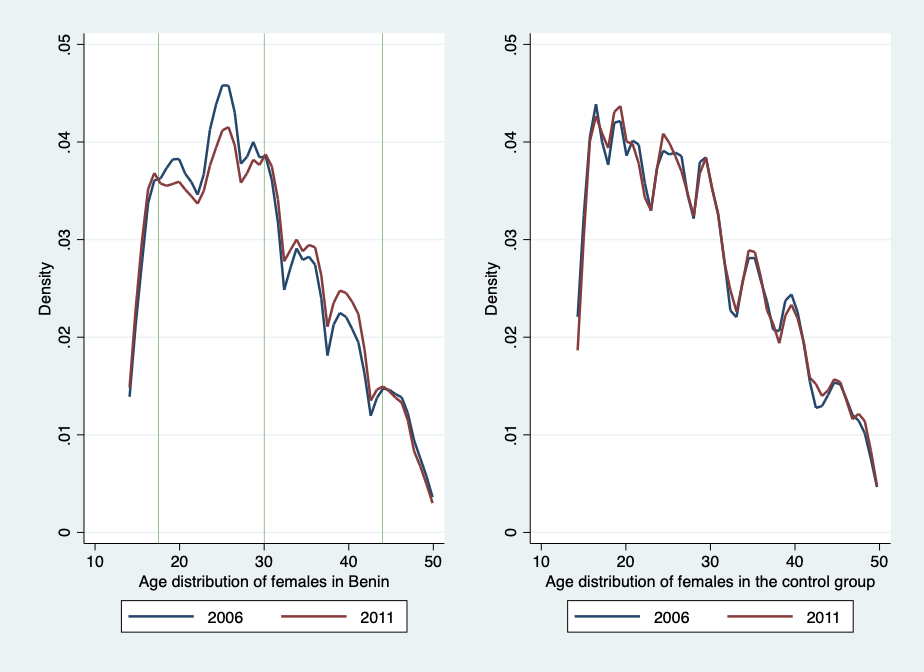}
    
 \uline{Source}: {DHS data} 
    \label{agedistribution}
\end{figure}

Figure \ref{csecuse} shows an increase in the use of cesarean section from less than 4 percent in 2001 and 2006 to 5.8 percent in 2011 and 5.9 percent in 2017. This represents an almost 50 percent increase after the implementation of the policy. Because of the exemption of the user fees, there are less financial constraints for people who can now use the cesarean section whenever it is possible. Though the procedure can prevent infant and maternal mortality, it could have a deadly effect on females if used too much or too soon, or without good expertise. I do not have information on the health status of females. Rather, I exploit the age distribution in the data. Figure \ref{fertilitybenin} shows the number of births per 1,000 women in Benin. Females aged between 25 and 30 are at the peak of fertility with more than 267 births per 1000 women. I assume that if the policy increases the mortality of females, the age distribution will change around 25-30 years. From figure \ref{agedistribution}, I see suggestive evidence that the policy has a positive effect on mortality. There are fewer females younger than 30 and more females older than 30 after the implementation of the policy in Benin. However, the distribution of age did not change in the control group. This suggests, maternal mortality might have gone up in Benin. I cannot directly measure maternal mortality from the DHS data. To verify my previous findings, I use the census data for Benin and the control group and construct a proxy for maternal death. Individuals were asked in the census if their mother was dead or alive. I restrict the data first to the youngest kid in each household and then, to kids who are five years old or younger. I assume that the younger the kid is, the more likely it is that their mother died due to childbearing. Therefore, I use the death status of the mother for young kids as a proxy for maternal mortality. For the most precise estimate, I restrict the data to only children who are younger than one year. The death of a female whose child is younger than one year is more likely due to recent childbearing than a kid who is five years old. However, my data only provides information on children who are alive. Therefore I face a selection issue as I cannot observe in the data what happened to the mothers whose children were dead during the delivery. To overcome this, I also use older kids in my regression. I assume that the mothers of the older kids might have tried to get a subsequent child and may be either dead or alive, and the child also may be dead or alive. Therefore, using older kids in this context allows me to recover several females that I would not have in my sample if I had used only younger kids. I acknowledge that my assumption is very strong, and therefore, my results should be taken with precaution.

The results in table \ref{mortmatsc0} suggest that access to the free cesarean section increases the probability of maternal mortality. Using kids less than five years old, the probability of maternal mortality went up by 5.21 percent\footnote{5.21 percentage change is obtained by computing the change from the mean of the dependent variable}. The result is smaller for kids less than one year with an increase of the likelihood of maternal death by 2.80 percent.  Thus, the policy effectively reduces infant mortality but, in some cases, at the expense of the mother's life.

\subsection{Family size}

Using different specifications, the Difference in Differences result presented in table \ref{fertility} suggests that having access to the exemption of Cesarean section user fees in Benin significantly decreases the total number of children a female has by 0.22. This represents a percentage change reduction of 7.48 percent of the number of children. I also notice that the more females are educated, the fewer children they have. In my findings, the age of females is positively correlated with the number of children, while their partner's age is negatively correlated with fertility. Marital status and the wealth index also significantly decrease the number of children a female has. 

The following mechanism can justify the negative relationship between access to the free cesarean section and fertility. \cite{ravit2018removing} found that access to the free cesarean section increases the use of the cesarean section in Benin and Mali, particularly in females who are non-educated, females who live in rural areas, and females in the middle-class wealth index. This result suggests that the monetary burden in Benin was minimized for women after the reform. Therefore if the cesarean section was needed, it was easier to use the surgery. However, once a female has undergone a cesarean section, there is a higher probability of repeating the process for subsequent pregnancies. After a cesarean section, the following pregnancy is associated with more risks, including the rupture of scar from the previous cesarean section, which can cause serious health problems and complications for a child. This may lead to high rates of repeated elective C section \citep{black2016planned}. In Benin, approximately 5 percent of females use cesarean section. Due to the related risks of subsequent deliveries, most doctors would advise women to wait, thereby allowing their bodies to recover sufficiently. This, therefore, explains why access to the free cesarean section would lower the ability and willingness of females to have more children. 

\begin{figure}[ht]
    \centering
    \caption{Distribution of Number of Children Females have}
        \label{Childdistribution}
    \includegraphics[width=.8\textwidth]{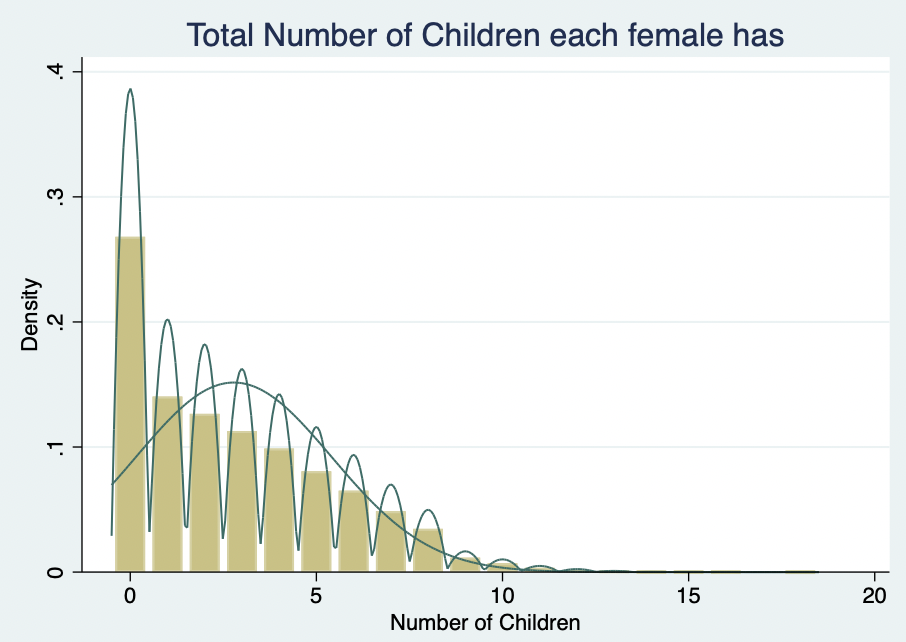}
    
 \uline{Source}: {DHS data} 
\end{figure}

To verify this argument, I exploit the distribution of the total number of children in the dataset (figure \ref{Childdistribution}) to analyze how the policy affects the fertility of females. I define binary variables indicating whether a female had a child at all, whether the female has a second child or not, and a binary variable indicating whether the female has more than two children. The table \ref{fertility51} summarises the results. The results suggest that access to the free cesarean section makes females less likely to have an additional child after the first baby. The probability of having a first child under the policy significantly increases by 3.36 percent while having a second child or more is less likely to happen. For females who had a child, the probability of having a second child decreased by $2.7$ percent. Furthermore, for a female who already had two children, the likelihood of having more children is even lower—the chance of having more than two children under the policy reduced by $3.65$ percent. These results suggest that, after the first delivery, females who underwent a cesarean section may have had health issues that prevented them from having a subsequent child.

\subsection{Maternal Labor Supply}

Another indirect effect of the cesarean section can be measured through maternal labor supply. Table \ref{labour1} describes the effect of the policy on females' labor supply. The results suggest that the policy has reduced the probability of the female working for up to three years following birth by 25.79 percent\footnote{I estimate the effect of the policy on the probability of maternal labor supply with a logit model. The results are the same as the linear probability model.}. 
Since fewer children correspond with reduced responsibilities in terms of childcare, I would have anticipated an increase in the labor supply after the implementation of the policy. However, I observe the opposite result here. The policy's negative and statistically significant effect on female labor supply may indicate adverse physical and mental health effects after the cesarean section, which does not permit females to return to their normal activities immediately. 

On the other hand, in Benin, a law protects Benin's pregnant women from being dismissed from their job because of a pregnancy. Pregnant women have the right to maternity leave which begins six weeks before the expected delivery date and ends eight weeks after delivery. During that leave, the mother deserves the full salary she was receiving at the suspension from work.
Also, during fifteen months from her resumption of work, the mother has the right to rest for breastfeeding for up to one hour per working day. These conditions may be non-beneficial for the employer and generate productivity loss due to the months of absence and reduced work hours. Therefore, the employer may dismiss a female who has not previously secured an indefinite work contract after her maternity period. More importantly, it may be the case that the eight weeks after delivery is not sufficient for the mother to heal completely and be able to go back to work. In this case, she would be choosing between caring for her newborn and herself and returning to her job at the expense of her health and her baby's health. The combination of sick paid leave and postnatal health problems is a possible justification for reducing the labor supply after birth. With free access to the cesarean section, females use more surgery for baby delivery. In turn, they reduce their labor supply to decrease the risk of stillbirth, miscarriage, and infant death.

\section{Discussion}
The main findings of my paper are that the presence of the free cesarean section in Benin significantly reduced infant mortality. The policy is most effective at reducing infant death for older females, females below the middle wealth index, and those with no education. Access to the policy also increases maternal mortality. In addition to the death of the female, it also might have some negative outcomes on the health of the females, which is translated by the reduction in fertility and maternal labor supply. Although the policy increases the probability of having a first child, it significantly decreases the likelihood of having a second child or even more than two children. Access to cesarean section can be associated with a lesser chance to conceive and a reduced probability of labor supply as an indirect effect. As childbearing costs fall when cesarean section becomes free, one would expect ceteris paribus the number of births to increase (which is the case for the first baby in my findings). Also, since fewer children are associated with reduced childcare responsibilities, one would expect that Maternal Labor Supply would increase. However, I have the opposite result in both cases. Thus, my view is that the policy might have harmed females' health, preventing them from immediately supplying labor or having more children than females who are not exposed to the policy.

Further work on this paper would investigate that plausible argument by evaluating the health outcomes of mothers who did a cesarean section compared to those who delivered naturally. Maternal mortality needs to be considered deeply with better data sources. Given the current data, I cannot say much about the health status of females because I lacked the necessary data on females' medical histories; the DHS survey is not panel data and is only conducted on females who are alive. However, this missing information might have affected my analyses by creating a selection problem. What I observed in the data is not the ideal population but the population that survives childbirth. This means that the population of mothers in Benin after the policy might differ from the population before the policy. The next step for improving the paper would be to look for a better data source to measure maternal mortality and maternal health outcomes. I provided evidence for maternal mortality of surviving children using the census data for Benin and my control group.

My recommendation, in that case, would be therefore, that first, the government continues to provide free access to the cesarean section. Second, they need to provide training to health professionals. This will avoid unnecessary baby deliveries by cesarean section and improve the quality of services offered. Third, they need to follow the sample of women who underwent cesarean sections (by monitoring the woman's and baby's health, injuries, controlling hemorrhages,  for up to 3 months, for example), etc.

\section{Conclusion}

The policy of free cesarean section was introduced in Benin in December 2008. Its implementation aims to prevent maternal death by improving financial accessibility to cesarean sections everywhere in the country. The removal of such a financial barrier positively impacts the take-up of cesarean sections, particularly on non-educated females, females who reside in rural zones, and females in the middle-class wealth index \citep{ravit2018removing}. 

This study looks at the impact of such accessibility first on infant and maternal mortality and second on the family size decision of females and the maternal labor supply decision after delivery. First, I find that access to Benin's free cesarean section significantly reduces the likelihood of infant mortality globally by a 36.85 percentage change. The policy is more effective at reducing infant death for older females, females who are below the middle wealth index, and females who have no education.  Second, access to free cesarean section increases the probability of maternal mortality.  Using kids less than five years old, the probability of maternal mortality went up by 5.21 percentage change with a 2.80 percentage change in the maternal mortality of kids less than one year. Third, the policy has negatively affected fertility with a percentage change reduction by 7.48 percent of the number of children. Even though the probability of having a first child significantly increases by 3.36 percentage change, the likelihood of having a second child or more than two children significantly decreases. This result suggests that after using the cesarean section for the first baby, females suffer from severe health issues that affect subsequent childbearing likelihood. Finally, having access to the cesarean section fee exemption policy reduced the probability of the female working for up to three years following birth by 25.79 percent. This result can be partly explained by the adverse health and emotional state effects on the females. The employment protection those females have from the law makes it easy for them to quit their job without any sanction or severance pay to their employer. My main recommendation is that the government, in addition to the free cesarean section, should make sure that women receive quality care during their pregnancy period, during the baby's birth, and after childbirth to save women's lives and the newborn. 


\newpage
\bibliographystyle{apalike}
\bibliography{myreference}


\newpage
\section*{Appendix}

%

\begin{figure}[ht]
    \centering
    \caption{Average number of children before and after the policy}
    \includegraphics[width=.8\textwidth]{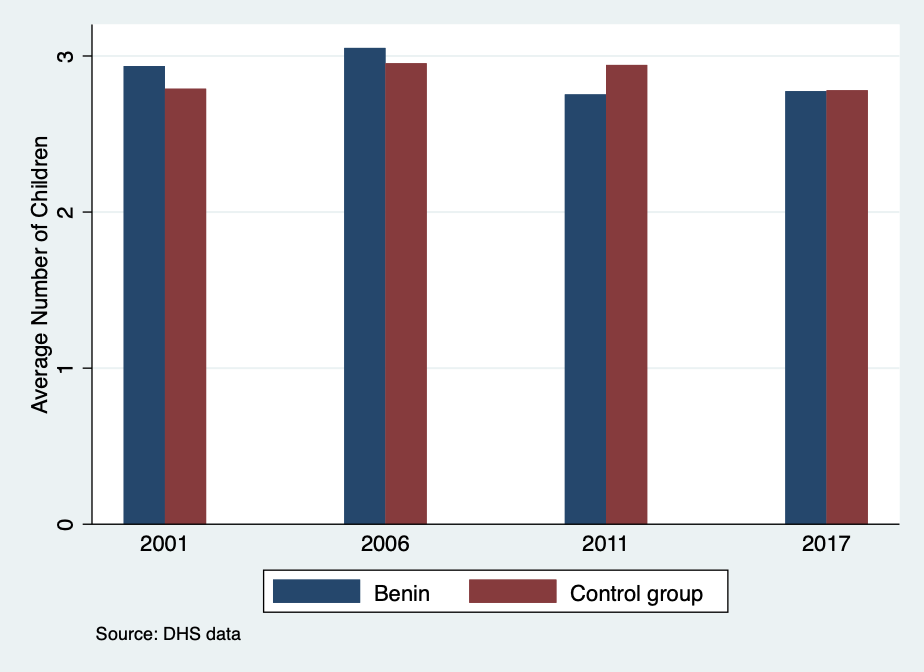}

    \label{kids}
\end{figure}


\begin{table}
\scriptsize
\centering
{
\def\sym#1{\ifmmode^{#1}\else\(^{#1}\)\fi}
\caption{Descriptive statistics \label{stat}}
\resizebox{\textwidth}{!}{
\begin{tabular}{l*{7}{c}}
\toprule
            &\multicolumn{3}{c}{Before Policy}&&\multicolumn{3}{c}{After Policy}\\
            &\multicolumn{1}{c}{Full sample}&\multicolumn{1}{c}{Treatment}&\multicolumn{1}{c}{Control}&&\multicolumn{1}{c}{Full sample}&\multicolumn{1}{c}{Treatment}&\multicolumn{1}{c}{Control}\\

\midrule
{\textbf{\uline{Age Group}}}&&&&&&&\\
15 - 24     &        0.41         &        0.37         &        0.42        & &        0.39         &        0.37         &        0.39         \\
            &     (0.492)         &     (0.482)         &     (0.494)       &  &     (0.487)         &     (0.484)         &     (0.488)         \\
\addlinespace
25 - 34     &        0.33         &        0.36         &        0.32       &  &        0.33         &        0.34         &        0.33         \\
            &     (0.470)         &     (0.480)         &     (0.467)       &  &     (0.470)         &     (0.475)         &     (0.469)         \\
\addlinespace
35 - 44     &        0.20         &        0.21         &        0.20       &  &        0.21         &        0.22         &        0.21         \\
            &     (0.398)         &     (0.406)         &     (0.396)       &  &     (0.409)         &     (0.411)         &     (0.409)         \\
\addlinespace
45 - 49     &        0.06         &        0.06         &        0.06       &  &        0.07         &        0.07         &        0.07         \\
            &     (0.239)         &     (0.243)         &     (0.237)       &  &     (0.255)         &     (0.250)         &     (0.256)         \\
{\textbf{\uline{Marital Status}}}&&&&&&\\
\addlinespace
Never in Union&        0.23         &        0.20         &        0.23       &  &        0.25         &        0.24         &        0.26         \\
            &     (0.418)         &     (0.403)         &     (0.421)       &  &     (0.434)         &     (0.428)         &     (0.436)         \\
\addlinespace
Married     &        0.67         &        0.66         &        0.67       &  &        0.61         &        0.55         &        0.62         \\
            &     (0.472)         &     (0.472)         &     (0.472)       &  &     (0.488)         &     (0.497)         &     (0.485)         \\
\addlinespace
With Partner&        0.06         &        0.08         &        0.05      &   &        0.08         &        0.15         &        0.07         \\
            &     (0.235)         &     (0.277)         &     (0.222)       &  &     (0.273)         &     (0.358)         &     (0.250)         \\
\addlinespace
Divorced    &        0.02         &        0.02         &        0.02       &  &        0.02         &        0.02         &        0.02         \\
            &     (0.141)         &     (0.137)         &     (0.142)       &  &     (0.149)         &     (0.134)         &     (0.152)         \\
\addlinespace
Widowed     &        0.01         &        0.01         &        0.01      &   &        0.01         &        0.01         &        0.01         \\
            &     (0.109)         &    (0.0875)         &     (0.114)      &   &     (0.107)         &    (0.0925)         &     (0.110)         \\
\addlinespace
Separated   &        0.02         &        0.02         &        0.02       &  &        0.02         &        0.03         &        0.02         \\
            &     (0.135)         &     (0.140)         &     (0.133)       &  &     (0.145)         &     (0.160)         &     (0.142)         \\
{\textbf{\uline{Wealth Index}}}&&&&&&\\
Poorest     &        0.19         &        0.19         &        0.19       &  &        0.17         &        0.18         &        0.17         \\
            &     (0.393)         &     (0.390)         &     (0.394)       &  &     (0.378)         &     (0.386)         &     (0.376)         \\
\addlinespace
Poorer      &        0.19         &        0.19         &        0.19       &  &        0.19         &        0.19         &        0.19         \\
            &     (0.389)         &     (0.389)         &     (0.389)       &  &     (0.392)         &     (0.393)         &     (0.391)         \\
\addlinespace
Middle      &        0.20         &        0.19         &        0.20       &  &        0.20         &        0.20         &        0.20         \\
            &     (0.399)         &     (0.395)         &     (0.400)       &  &     (0.402)         &     (0.398)         &     (0.402)         \\
\addlinespace
Richer      &        0.20         &        0.21         &        0.19       &  &        0.21         &        0.21         &        0.21         \\
            &     (0.398)         &     (0.408)         &     (0.395)       &  &     (0.408)         &     (0.407)         &     (0.409)         \\
\addlinespace
Richest     &        0.23         &        0.22         &        0.23       &  &        0.22         &        0.22         &        0.22         \\
            &     (0.419)         &     (0.416)         &     (0.420)       &  &     (0.417)         &     (0.414)         &     (0.418)         \\
\multicolumn{2}{c}{\textbf{\uline{Use  of  Cesarean Section  }}}\\
\addlinespace
No          &        0.98         &        0.96         &        0.98       &  &        0.96         &        0.94         &        0.97         \\
            &     (0.146)         &     (0.192)         &     (0.131)       &  &     (0.186)         &     (0.233)         &     (0.174)         \\
\addlinespace
Yes         &        0.02         &        0.04         &        0.02       &  &        0.04         &        0.06         &        0.03         \\
            &     (0.146)         &     (0.192)         &     (0.131)       &  &     (0.186)         &     (0.233)         &     (0.174)         \\
\midrule
Observations       &      115218         &       23323         &       91895       &  &      185818         &       32083         &      153735         \\
\bottomrule
\multicolumn{8}{l}{\footnotesize Mean coefficients are reported in the table; standard deviation are in parentheses. Treatment here represents}\\
\multicolumn{8}{l}{\footnotesize Benin females. Control represents females in Burkina Faso, Cameroon, Cote d'Ivoire, Niger, Nigeria and Togo.}\\
\end{tabular}
}
}
\end{table}

\begin{center}
\begin{table}[ht]
    \centering
\caption{Number of infant death and stillbirth}
\label{mortinfant11}
\begin{tabular}{l*{4}{c}}
\toprule
            &\multicolumn{4}{c}{Infant Mortality (number of infant death/stillbirth)}\\
            &\multicolumn{1}{c}{(1)}&\multicolumn{1}{c}{(2)}&\multicolumn{1}{c}{(3)}&\multicolumn{1}{c}{(4)}\\
\midrule
Treatment   &      -0.0734&      -0.0754&      -0.0884&      -0.0855\\
            &    (0.0280)&    (0.0281)&    (0.0293)&    (0.0290)\\
\addlinespace
Education Attainment&      -0.118&      -0.119&     -0.0848&     -0.0869\\
            &    (0.0144)&    (0.0145)&    (0.0128)&    (0.0131)\\
\addlinespace
Age of Female&      0.0359&      0.0355&      0.0359&      0.0372\\
            &   (0.00345)&   (0.00344)&   (0.00336)&   (0.00356)\\
\addlinespace
Age of Partner&            &     0.00137&     0.00131&    0.000928\\
            &            &  (0.000350)&  (0.000325)&  (0.000277)\\
\addlinespace
Wealth Index&            &            &     -0.0770&     -0.0771\\
            &            &            &    (0.0127)&    (0.0126)\\
\addlinespace
Marital Status&            &            &            &     -0.0385\\
            &            &            &            &   (0.00965)\\
\addlinespace
Country Fixed Effect &   Yes   &  Yes      & Yes      &    Yes\\
Time Fixed Effect.  &   Yes   &  Yes      & Yes      &    Yes\\
\addlinespace
\_cons      &      -0.388&      -0.438&      -0.250&      -0.231\\
            &    (0.0851)&    (0.0793)&    (0.0708)&    (0.0674)\\
\midrule
Observations&      196554&      196424&      196424&      196423\\
\% Change &         -0.1609&         -0.1657&  -0.1942&     -0.1879\\
R-squared   &       0.182&       0.182&       0.192&       0.194\\
\midrule
Mean Dep. Var.&       0.456&       0.455&       0.455&       0.455\\
\bottomrule
\multicolumn{5}{l}{\footnotesize Standard errors in parentheses; Standard errors are clustered at the household level.}\\
\multicolumn{5}{l}{\footnotesize Dep. Var. Mean is the dependent variable mean. Treatment is Benin Females after 2008}\\
\multicolumn{5}{l}{\footnotesize \% Change = $\hat{\beta_1} / \Bar{Y_i}$}\\
\end{tabular}
\end{table}
\end{center}

\begin{center}
\begin{table}[h!]
    \centering
\caption{Probability of infant death and stillbirth \label{mortinfant21}}
\begin{tabular}{l*{4}{c}}
\toprule
            &\multicolumn{4}{c}{Probability of Infant death or stillbirth}\\
            &\multicolumn{1}{c}{(1)}&\multicolumn{1}{c}{(2)}&\multicolumn{1}{c}{(3)}&\multicolumn{1}{c}{(4)}\\
\midrule
Treatment   &     -0.0911&     -0.0908&     -0.0961&     -0.0962\\
            &    (0.0113)&    (0.0113)&    (0.0126)&    (0.0129)\\
\addlinespace
Education Attainment&     -0.0578&     -0.0576&     -0.0443&     -0.0443\\
            &   (0.00346)&   (0.00340)&   (0.00415)&   (0.00420)\\
\addlinespace
Age of Female&      0.0168&      0.0170&      0.0171&      0.0171\\
            &   (0.00108)&   (0.00112)&   (0.00104)&   (0.00108)\\
\addlinespace
Age of Partner&            &   -0.000425&   -0.000448&   -0.000435\\
            &            &  (0.000183)&  (0.000151)&  (0.000152)\\
\addlinespace
Wealth Index&            &            &     -0.0302&     -0.0302\\
            &            &            &   (0.00277)&   (0.00277)\\
\addlinespace
Marital Status&            &            &            &     0.00128\\
            &            &            &            &   (0.00421)\\
\addlinespace
Country Fixed Effect &   Yes   &  Yes      & Yes      &    Yes\\
\addlinespace
Time Fixed Effect.  &   Yes   &  Yes      & Yes      &    Yes\\
\addlinespace
\_cons      &      -0.129&      -0.114&     -0.0400&     -0.0406\\
            &    (0.0323)&    (0.0298)&    (0.0266)&    (0.0255)\\
\midrule
Observations&      196554&      196424&      196424&      196423\\

\% Change &         -0.3490&      -0.3478&  -0.3681&       -0.3685\\
R-squared   &       0.196&       0.197&       0.204&       0.204\\
\midrule
Mean Dep. Var.&       0.261&       0.261&       0.261&       0.261\\
\bottomrule
\multicolumn{5}{l}{\footnotesize Standard errors in parentheses; Standard errors are clustered at the household level.}\\
\multicolumn{5}{l}{\footnotesize Dep. Var. Mean is the dependent variable mean. Treatment is Benin Females after 2008}\\
\multicolumn{5}{l}{\footnotesize \% Change = $\hat{\beta_1} / \Bar{Y_i}$}\\
\end{tabular}
\end{table}
\end{center}

\begin{center}
\begin{table}[h!]
    \centering
\caption{Probability of infant mortality by age group, wealth index and Education level}  
\label{mortinfant31}
\begin{tabular}{l*{7}{c}}
\toprule
    \textbf{Age Group}       &\multicolumn{1}{c}{}&\multicolumn{1}{c}{}&\multicolumn{1}{c}{15-24}&\multicolumn{1}{c}{25-34}&\multicolumn{1}{c}{35-44}&\multicolumn{1}{c}{45-49}\\
\midrule
Treatment   & &&   -0.00337&     -0.0941&      -0.202&      -0.241\\
            & &&  (0.00583)&    (0.0201)&    (0.0217)&    (0.0130)\\
\addlinespace
\midrule
Observations&  & &      76837&       65571&       40982&       13033\\
\% Change  &  & &        -0.0465&  -0.3085&   -0.4353&      -0.4725\\
Mean Dep. Var.&  & &     0.0724&       0.305&       0.464&       0.510\\
\midrule
    \textbf{Wealth Index}       &\multicolumn{1}{c}{}&\multicolumn{1}{c}{Poorest}&\multicolumn{1}{c}{Poorer}&\multicolumn{1}{c}{Middle}&\multicolumn{1}{c}{Richer}&\multicolumn{1}{c}{Richest}\\
\midrule
Treatment   &     & -0.172&      -0.119&      -0.116&     -0.0636&     -0.0224\\
           & &    (0.0202)&    (0.0148)&    (0.0100)&    (0.0144)&    (0.0127)\\
\addlinespace
\midrule
Observations&      & 35497&       36590&       38869&       40297&       45170\\
\% Change  &      & -0.4636&    -0.3584&      -0.4172&     -0.2852&      -0.1647\\
Mean Dep. Var.&   &    0.371&       0.332&       0.278&       0.223&       0.136\\
\midrule
    \textbf{Education Attainment}         &\multicolumn{1}{c}{0}&\multicolumn{1}{c}{P1}&\multicolumn{1}{c}{P2}&\multicolumn{1}{c}{S1}&\multicolumn{1}{c}{S2}&\multicolumn{1}{c}{H}\\
\midrule
Treatment   &      -0.143&     -0.0504&     -0.0462&     0.00283&      0.00157&     -0.00110\\
            &    (0.0174)&    (0.0178)&    (0.0147)&   (0.00823)&   (0.00202)&   (0.00498)\\
\midrule
Observations&       96324&       23301&       17345&       37318&       13664&        8442\\
\% Change  &       -0.3907&     -0.2040& -0.1811&   -0.0291&    -0.0138 &     -0.014\\
Mean Dep. Var.&       0.366&       0.247&       0.255&      0.0971&       0.113&      0.0782\\
\bottomrule
\multicolumn{7}{l}{\footnotesize Education Attainment 0 = No education, P1 = Incomplete Primary, P2 = Complete Primary, }\\
\multicolumn{7}{l}{\footnotesize  S1 = Incomplete Secondary, S2 = Complete Secondary, H = Higher Education }\\
\multicolumn{7}{l}{\footnotesize  Standard errors are clustered at the household level. Dep. Var. Mean represents the dependent variable mean.}\\
\multicolumn{7}{l}{\footnotesize  Country and time fixed effects are included. Treatment is Benin Females after 2008}\\
\multicolumn{7}{l}{\footnotesize \% Change = $\hat{\beta_1} / \Bar{Y_i}$. Estimates in this table are extracted from tables \ref{mortinfant3} \ref{mortinfant4} \ref{mortinfant5} in the appendix. }\\
\end{tabular}
\end{table}
\end{center}

\begin{center}
\begin{table}[h!]
    \centering

\caption{Maternal Mortality of Surviving Children \label{mortmatsc0}}
\begin{tabular}{l*{6}{c}}
\toprule
            &\multicolumn{6}{c}{\uline{Mortality status of mother: 1 if dead and 0 if alive}}\\
            \textbf{Age of Child}          &\multicolumn{1}{c}{5 or less}&\multicolumn{1}{c}{4 or less}&\multicolumn{1}{c}{3 or less}&\multicolumn{1}{c}{2 or less}&\multicolumn{1}{c}{1 or less}&\multicolumn{1}{c}{$<$ 1 year}\\
\midrule
Treatment   &     0.00465&     0.00392&     0.00330&     0.00225&     0.00201&     0.00123\\
            &  (0.000482)&  (0.000465)&  (0.000449)&  (0.000425)&  (0.000461)&  (0.000595)\\
\addlinespace
\midrule
Observations&     920747&      860608&      784798&      664633&      472571&      262212\\
\% Change &     0.0521&         0.0502&         0.0494&         0.0408&         0.0419&         0.0280\\
\midrule
Mean Dep. Var.&     0.0891&     0.0780&     0.0668&     0.0551&     0.0479&     0.0438\\
\bottomrule
\multicolumn{7}{l}{\footnotesize Standard errors in parentheses; Demographic characteristics variables include sex and region}\\
\multicolumn{7}{l}{\footnotesize Country and time fixed effects are included and standard errors are clustered at the household level.}\\
\multicolumn{7}{l}{\footnotesize Dep. Var. Mean represents the dependent variable mean.}\\
\multicolumn{7}{l}{\footnotesize Treatment is Benin Females after 2008. Estimates in this table are extracted from tables \ref{mortmatsc1} in the appendix.}\\
\end{tabular}
\end{table}
\end{center}

\begin{table}[h!]
    \centering

\caption{Effect of free cesarean section on fertility \label{fertility}}
\begin{tabular}{l*{4}{c}}
\toprule
            &\multicolumn{4}{c}{Number of children}\\
            &\multicolumn{1}{c}{(1)}&\multicolumn{1}{c}{(2)}&\multicolumn{1}{c}{(3)}&\multicolumn{1}{c}{(4)}\\
\midrule
Treatment   &      -0.198&      -0.191&      -0.226&      -0.220\\
            &    (0.0203)&    (0.0203)&    (0.0200)&    (0.0200)\\
\addlinespace
Education Attainment&      -0.431&      -0.426&      -0.339&      -0.343\\
            &   (0.00237)&   (0.00239)&   (0.00273)&   (0.00276)\\
\addlinespace
Age of Female&       0.207&       0.210&       0.211&       0.213\\
            &  (0.000511)&  (0.000502)&  (0.000499)&  (0.000544)\\
\addlinespace
Age of Partner&            &    -0.00880&    -0.00894&    -0.00958\\
            &            &  (0.000253)&  (0.000252)&  (0.000256)\\
\addlinespace
Wealth Index&            &            &      -0.197&      -0.198\\
            &            &            &   (0.00319)&   (0.00319)\\
\addlinespace
Marital Status&            &            &            &     -0.0644\\
            &            &            &            &   (0.00516)\\
\addlinespace
Country Fixed Effect &   Yes   &  Yes      & Yes      &    Yes\\
Time Fixed Effect.  &   Yes   &  Yes      & Yes      &    Yes\\
\addlinespace
\_cons      &      -2.321&      -1.998&      -1.518&      -1.486\\
            &    (0.0134)&    (0.0174)&    (0.0188)&    (0.0189)\\
\midrule
Observations&      196554&      196424&      196424&      196423\\
\% Change&        -0.0673&      -0.0649&   -0.0768&    -0.0748\\
R-squared   &       0.591&       0.593&       0.601&       0.602\\
\midrule
Mean Dep. Var.&       2.940&       2.940&       2.940&       2.940\\
\bottomrule
\multicolumn{5}{l}{\footnotesize Standard errors in parentheses and clustered at the household level.}\\
\multicolumn{5}{l}{\footnotesize Dep. Var. Mean is the dependent variable mean. Treatment is Benin Females after 2008}\\
\end{tabular}
\end{table}

\begin{center}
    \begin{table}[h!]
        \centering
\caption{Marginal effects of free cesarean section \label{fertility51}}
\begin{tabular}{l*{4}{c}}
\toprule
            &\multicolumn{1}{c}{Having } &\multicolumn{1}{c}{ }&\multicolumn{1}{c}{Having}&\multicolumn{1}{c}{Having}\\
            &\multicolumn{1}{c}{a first child} &\multicolumn{1}{c}{ }&\multicolumn{1}{c}{a second child}&\multicolumn{1}{c}{more than 2 children}\\
\midrule
Treatment   &     0.0247&    &     -0.0164&     -0.0178\\
            &   (0.00367)&  & (0.00418)&   (0.00443)\\
\midrule
Observations&      196423&    &  196423&      196423\\
\% Change &       0.03369&    &  -0.02706&     -0.0365\\
R-squared   &       0.468&    &   0.480&       0.478\\
\midrule
Mean Dep. Var.&       0.733&   &    0.606&       0.487\\
\bottomrule
\multicolumn{5}{l}{\footnotesize Standard errors in parentheses; Standard errors are clustered at the household level. Dep. Var. Mean }\\
\multicolumn{5}{l}{\footnotesize is the dependent variable mean. Estimates in this table are extracted from tables \ref{fertility2}, \ref{fertility3} and \ref{fertility4}.}\\
\end{tabular}

    \end{table}
\end{center}

\begin{table}[h!]
    \centering
\caption{Effects of the access to free cesarean section on maternal labor supply \label{labour1}}
\begin{tabular}{l*{5}{c}}
\toprule
            &\multicolumn{5}{c}{Maternal Labor Supply: 1 if working, 0 otherwise}\\
            &\multicolumn{1}{c}{(1)}&\multicolumn{1}{c}{(2)}&\multicolumn{1}{c}{(3)}&\multicolumn{1}{c}{(4)}&\multicolumn{1}{c}{(5)}\\
\midrule
Treatment   &      -0.158&      -0.156&      -0.158&      -0.161&      -0.163\\
            &   (0.00522)&   (0.00520)&   (0.00520)&   (0.00519)&   (0.00520)\\
\addlinespace
Education Attainment&     -0.0102&    -0.00890&    -0.00409&    -0.00221&    -0.00408\\
            &  (0.000664)&  (0.000664)&  (0.000761)&  (0.000763)&  (0.000789)\\
\addlinespace
Age of Female&      0.0157&      0.0163&      0.0163&      0.0151&      0.0162\\
            &  (0.000109)&  (0.000110)&  (0.000111)&  (0.000120)&  (0.000170)\\
\addlinespace
Age of Partner&            &    -0.00218&    -0.00219&    -0.00184&    -0.00189\\
            &            & (0.0000773)& (0.0000773)& (0.0000782)& (0.0000783)\\
\addlinespace
Wealth Index&            &            &     -0.0110&     -0.0109&     -0.0120\\
            &            &            &  (0.000841)&  (0.000840)&  (0.000851)\\
\addlinespace
Marital Status&            &            &            &      0.0351&      0.0348\\
            &            &            &            &   (0.00122)&   (0.00122)\\
\addlinespace
Number of Children    &            &            &            &            &    -0.00546\\
            &            &            &            &            &  (0.000588)\\
\addlinespace
Country Fixed Effect &   Yes   &  Yes      & Yes      &    Yes& Yes\\
\addlinespace
Time Fixed Effect  &   Yes   &  Yes      & Yes      &    Yes& Yes\\
\addlinespace
\_cons      &       0.215&       0.295&       0.321&       0.304&       0.296\\
            &   (0.00363)&   (0.00459)&   (0.00501)&   (0.00504)&   (0.00509)\\
\midrule
Observations&      196215&      196085&      196085&      196084&      196084\\
\% Change&        -0.25&       -0.2468&       -0.25&       -0.2547&       -0.2579\\
R-squared   &       0.104&       0.108&       0.109&       0.112&       0.113\\
\midrule
Mean Dep. Var.&       0.632&       0.632&       0.632&       0.632&       0.632\\
\bottomrule
\multicolumn{6}{l}{\footnotesize Standard errors in parentheses and clustered at the household level.}\\
\multicolumn{6}{l}{\footnotesize Dep. Var. Mean represents the dependent variable mean. Treatment is Benin Females after 2008}\\
\end{tabular}
    \end{table}

\begin{center}
\begin{table}[h!]
    \centering

\caption{Probability of infant death and stillbirth by age group \label{mortinfant3}}
\begin{tabular}{l*{4}{c}}
\toprule
            &\multicolumn{1}{c}{Age 15-24}&\multicolumn{1}{c}{Age 25-34}&\multicolumn{1}{c}{Age 35-44}&\multicolumn{1}{c}{Age 45-49}\\
\midrule
Treatment   &    -0.00337&     -0.0941&      -0.202&      -0.241\\
            &   (0.00583)&    (0.0201)&    (0.0217)&    (0.0130)\\
\addlinespace
Education Attainment&     -0.0197&     -0.0570&     -0.0605&     -0.0533\\
            &   (0.00323)&   (0.00498)&   (0.00925)&    (0.0132)\\
\addlinespace
Age of Female&      0.0174&      0.0193&     0.00690&     0.00428\\
            &   (0.00174)&   (0.00159)&  (0.000870)&   (0.00308)\\
\addlinespace
Age of Partner&   -0.000650&   -0.000152&    0.000465&    0.000276\\
            & (0.0000909)&  (0.000327)&  (0.000383)&  (0.000594)\\
\addlinespace
Wealth Index&     -0.0117&     -0.0383&     -0.0446&     -0.0359\\
            &   (0.00104)&   (0.00429)&   (0.00599)&   (0.00881)\\
\addlinespace
Marital Status&      0.0190&     0.00477&     -0.0130&    -0.00552\\
            &   (0.00246)&   (0.00735)&   (0.00565)&   (0.00373)\\
\addlinespace
Country Fixed Effect &   Yes   &  Yes      & Yes      &    Yes\\
\addlinespace
Time Fixed Effect.  &   Yes   &  Yes      & Yes      &    Yes\\
\addlinespace
\_cons      &      -0.174&     -0.0507&       0.410&       0.480\\
            &    (0.0279)&    (0.0328)&    (0.0491)&     (0.142)\\
\midrule
Observations&       76837&       65571&       40982&       13033\\
R-squared   &      0.0888&       0.102&      0.0969&      0.0687\\
\midrule
Mean Dep. Var.&      0.0724&       0.305&       0.464&       0.510\\
\bottomrule
\multicolumn{5}{l}{\footnotesize Standard errors in parentheses; Standard errors are clustered at the household level.}\\
\multicolumn{5}{l}{\footnotesize Dep. Var. Mean is the dependent variable mean. Treatment is Benin Females after 2008}\\
\end{tabular}

\end{table}
\end{center}

\begin{center}
\begin{table}[h!]
    \centering
\caption{Probability of infant death and stillbirth by the Wealth Index of the female \label{mortinfant4}}
\begin{tabular}{l*{5}{c}}
\toprule
            &\multicolumn{1}{c}{Poorest}&\multicolumn{1}{c}{Poorer}&\multicolumn{1}{c}{Middle}&\multicolumn{1}{c}{Richer}&\multicolumn{1}{c}{Richest}\\
\midrule
Treatment   &      -0.172&      -0.119&      -0.116&     -0.0636&     -0.0224\\
            &    (0.0202)&    (0.0148)&    (0.0100)&    (0.0144)&    (0.0127)\\
\addlinespace
Education Attainment&     -0.0335&     -0.0470&     -0.0420&     -0.0465&     -0.0394\\
            &   (0.00595)&   (0.00379)&   (0.00493)&   (0.00609)&   (0.00492)\\
\addlinespace
Age of Female&      0.0219&      0.0196&      0.0174&      0.0159&      0.0118\\
            &  (0.000960)&   (0.00117)&   (0.00148)&   (0.00137)&   (0.00134)\\
\addlinespace
Age of Partner&    -0.00141&   -0.000957&   -0.000827&   -0.000181&    0.000296\\
            &  (0.000143)&  (0.000215)&  (0.000182)&  (0.000250)&  (0.000215)\\
\addlinespace
Marital Status&     0.00351&   0.0000309&     0.00540&     0.00181&     0.00851\\
            &   (0.00537)&   (0.00322)&   (0.00688)&   (0.00819)&   (0.00389)\\
\addlinespace
Country Fixed Effect &   Yes   &  Yes      & Yes      &    Yes&Yes\\
\addlinespace
Time Fixed Effect.  &   Yes   &  Yes      & Yes      &    Yes&Yes\\
\addlinespace
\_cons      &      -0.177&      -0.140&      -0.123&      -0.129&      -0.114\\
            &    (0.0255)&    (0.0261)&    (0.0292)&    (0.0234)&    (0.0251)\\
\midrule
Observations&       35497&       36590&       38869&       40297&       45170\\
Fixed Efects&         Yes&         Yes&         Yes&         Yes&         Yes\\
R-squared   &       0.191&       0.186&       0.180&       0.182&       0.151\\
\midrule
Mean Dep. Var.&       0.371&       0.332&       0.278&       0.223&       0.136\\
\bottomrule
\multicolumn{6}{l}{\footnotesize Standard errors in parentheses; Standard errors are clustered at the household level.}\\
\multicolumn{6}{l}{\footnotesize Dep. Var. Mean represents the dependent variable mean. Treatment is Benin Females after 2008}\\
\end{tabular}

\end{table}
\end{center}

\begin{center}
\begin{table}[h!]
    \centering
\caption{Probability of infant death and stillbirth by Education attainment of the female \label{mortinfant5}}
\begin{tabular}{l*{6}{c}}
\toprule
            &\multicolumn{1}{c}{0}&\multicolumn{1}{c}{P1}&\multicolumn{1}{c}{P2}&\multicolumn{1}{c}{S1}&\multicolumn{1}{c}{S2}&\multicolumn{1}{c}{H}\\
\midrule
Treatment   &      -0.143&     -0.0504&     -0.0462&     0.00283&      0.00157&     -0.00110\\
            &    (0.0174)&    (0.0178)&    (0.0147)&   (0.00823)&   (0.00202)&   (0.00498)\\
\addlinespace
Age of Female&      0.0199&      0.0181&      0.0152&      0.0124&      0.0116&     0.00926\\
            &   (0.00141)&  (0.000991)&   (0.00114)&   (0.00143)&  (0.000551)&  (0.000594)\\
\addlinespace
Age of Partner&   -0.000929&   -0.000731&    -0.00102&   -0.000578&   -0.000289&   0.0000401\\
            &  (0.000319)&  (0.000250)&  (0.000208)&  (0.000148)& (0.0000405)& (0.0000669)\\
\addlinespace
Wealth Index&     -0.0282&     -0.0362&     -0.0383&     -0.0241&     -0.0217&     -0.0371\\
            &   (0.00491)&   (0.00223)&   (0.00498)&   (0.00167)&   (0.00203)&   (0.00178)\\
\addlinespace
Marital Status&    -0.00105&     0.00618&     0.00984&      0.0109&      0.0251&      0.0218\\
            &   (0.00542)&   (0.00244)&   (0.00437)&   (0.00475)&   (0.00322)&   (0.00740)\\
\addlinespace
Country Fixed Effect &   Yes   &  Yes      & Yes      &    Yes&Yes&Yes\\
\addlinespace
Time Fixed Effect.  &   Yes   &  Yes      & Yes      &    Yes&Yes&Yes\\
\addlinespace
\_cons      &      -0.102&      -0.111&     -0.0336&     -0.0872&      -0.119&     -0.0471\\
            &    (0.0339)&    (0.0199)&    (0.0129)&    (0.0330)&    (0.0194)&    (0.0263)\\
\midrule
Observations&       96324&       23301&       17345&       37318&       13664&        8442\\
Fixed Efects&         Yes&         Yes&         Yes&         Yes&         Yes&         Yes\\
R-squared   &       0.156&       0.175&       0.119&       0.152&       0.110&      0.0996\\
\midrule
Mean Dep. Var.&       0.366&       0.247&       0.255&      0.0971&       0.113&      0.0782\\
\bottomrule
\multicolumn{7}{l}{\footnotesize 0 = No education, P1 = Incomplete Primary, P2 = Complete Primary, S1 = Incomplete Secondary,}\\
\multicolumn{7}{l}{\footnotesize  S2 = Complete Secondary, H = Higher. Standard errors in parentheses and clustered at the household level.}\\
\multicolumn{7}{l}{\footnotesize Dep. Var. Mean represents the dependent variable mean. Treatment is Benin Females after 2008}\\
\end{tabular}
\end{table}
\end{center}

\begin{center}
\begin{table}[h!]
    \centering

\caption{Maternal Mortality of Surviving Children \label{mortmatsc1}}
\begin{tabular}{l*{6}{c}}
\toprule
            &&\multicolumn{5}{l}{Mortality status of mother: 1 if dead and 0 if alive }\\
            \textbf{Age of Child}          &\multicolumn{1}{c}{5 or less}&\multicolumn{1}{c}{4 or less}&\multicolumn{1}{c}{3 or less}&\multicolumn{1}{c}{2 or less}&\multicolumn{1}{c}{1 or less}&\multicolumn{1}{c}{$<$ 1 year}\\
\midrule
Treatment   &     0.00465&     0.00392&     0.00330&     0.00225&     0.00201&     0.00123\\
            &  (0.000482)&  (0.000465)&  (0.000449)&  (0.000425)&  (0.000461)&  (0.000595)\\
\addlinespace
\addlinespace
Gender of child: &    0.000354&    0.000257&    0.000331&    0.000324&    0.000340&    0.000141\\
 1 for female           &  (0.000196)&  (0.000190)&  (0.000184)&  (0.000182)&  (0.000201)&  (0.000258)\\
\addlinespace
\addlinespace
Type of residence:       &    -0.00117&   -0.000715&   -0.000400&  -0.0000565&  -0.0000571&    0.000333\\
  1 for urban           &  (0.000202)&  (0.000198)&  (0.000193)&  (0.000192)&  (0.000212)&  (0.000277)\\
\addlinespace
\addlinespace
Country Fixed Effect &   Yes   &  Yes      & Yes      &    Yes&Yes&yes\\
\addlinespace
Time Fixed Effect.  &   Yes   &  Yes      & Yes      &    Yes&Yes&Yes\\
\addlinespace
\_cons      &     0.00871&     0.00755&     0.00634&     0.00515&     0.00444&     0.00406\\
            &  (0.000159)&  (0.000154)&  (0.000148)&  (0.000146)&  (0.000162)&  (0.000210)\\
\midrule
Observations&     920747&      860608&      784798&      664633&      472571&      262212\\
\% Change &     0.0521&         0.0502&         0.0494&         0.0408&         0.0419&         0.0280\\
R-squared   &    0.000626&    0.000539&    0.000411&    0.000358&    0.000300&    0.000267\\
\midrule
Mean Dep. Var.&     0.0891&     0.0780&     0.0668&     0.0551&     0.0479&     0.0438\\
\bottomrule
\multicolumn{7}{l}{\footnotesize Standard errors in parentheses; Demographic characteristics variables include gender and region}\\
\multicolumn{7}{l}{\footnotesize Dep. Var. Mean represents the dependent variable mean. Treatment is Benin Females after 2008.}\\
\end{tabular}
\end{table}
\end{center}

\begin{center}
\begin{table}[h!]
    \centering

\caption{Effect of free cesarean section on the probability of having a child \label{fertility2}}
\begin{tabular}{l*{4}{c}}
\toprule
            &\multicolumn{4}{c}{Having a first child: 1 if yes, 0 otherwise}\\            &\multicolumn{1}{c}{(1)}&\multicolumn{1}{c}{(2)}&\multicolumn{1}{c}{(3)}&\multicolumn{1}{c}{(4)}\\
\midrule
Treatment   &     0.0185&     0.0121&     0.0147&     0.0247\\
            &   (0.00405)&   (0.00384)&   (0.00383)&   (0.00367)\\
\addlinespace
Education Attainment&     -0.0680&     -0.0635&     -0.0571&     -0.0507\\
            &  (0.000554)&  (0.000539)&  (0.000610)&  (0.000581)\\
\addlinespace
Age of Female&      0.0250&      0.0269&      0.0270&      0.0228\\
            & (0.0000879)& (0.0000849)& (0.0000848)& (0.0000882)\\
\addlinespace
Age of Partner&            &    -0.00738&    -0.00739&    -0.00622\\
            &            & (0.0000640)& (0.0000639)& (0.0000614)\\
\addlinespace
Wealth Index&            &            &     -0.0146&     -0.0144\\
            &            &            &  (0.000607)&  (0.000580)\\
\addlinespace
Marital Status&            &            &            &       0.118\\
            &            &            &            &   (0.00120)\\
\addlinespace
Country Fixed Effect &   Yes   &  Yes      & Yes      &    Yes\\
\addlinespace
Time Fixed Effect.  &   Yes   &  Yes      & Yes      &    Yes\\
\addlinespace
\_cons      &       0.120&       0.389&       0.425&       0.365\\
            &   (0.00312)&   (0.00371)&   (0.00395)&   (0.00390)\\
\midrule
Observations&      196554&      196424&      196424&      196423\\
R-squared   &       0.367&       0.419&       0.421&       0.468\\
\midrule
Mean Dep. Var.&       0.733&       0.733&       0.733&       0.733\\
\bottomrule
\multicolumn{5}{l}{\footnotesize Standard errors in parentheses and clustered at the household level.}\\
\multicolumn{5}{l}{\footnotesize Dep. Var. Mean represents the dependent variable mean. Treatment is Benin Females after 2008}\\
\end{tabular}

\end{table}
\end{center}

\begin{center}
\begin{table}[h!]
    \centering
\caption{Effect of free cesarean section on the probability of having a second child \label{fertility3}}
\begin{tabular}{l*{4}{c}}
\toprule
            &\multicolumn{4}{c}{Having a second child: 1 if yes, 0 otherwise}\\                     &\multicolumn{1}{c}{(1)}&\multicolumn{1}{c}{(2)}&\multicolumn{1}{c}{(3)}&\multicolumn{1}{c}{(4)}\\
\midrule
Treatment   &     -0.0134&    -0.00870&     -0.0121&     -0.0164\\
            &   (0.00431)&   (0.00421)&   (0.00420)&   (0.00418)\\
\addlinespace
Education Attainment&     -0.0722&     -0.0689&     -0.0605&     -0.0578\\
            &  (0.000549)&  (0.000545)&  (0.000615)&  (0.000613)\\
\addlinespace
Age of Female&      0.0315&      0.0330&      0.0331&      0.0312\\
            & (0.0000845)& (0.0000812)& (0.0000812)& (0.0000930)\\
\addlinespace
Age of Partner&            &    -0.00544&    -0.00546&    -0.00496\\
            &            & (0.0000617)& (0.0000616)& (0.0000627)\\
\addlinespace
Wealth Index&            &            &     -0.0192&     -0.0191\\
            &            &            &  (0.000641)&  (0.000636)\\
\addlinespace
Marital Status&            &            &            &      0.0506\\
            &            &            &            &   (0.00125)\\
\addlinespace
Country Fixed Effect &   Yes   &  Yes      & Yes      &    Yes\\
\addlinespace
Time Fixed Effect.  &   Yes   &  Yes      & Yes      &    Yes\\
\addlinespace
\_cons      &      -0.186&      0.0126&      0.0592&      0.0338\\
            &   (0.00294)&   (0.00393)&   (0.00423)&   (0.00425)\\
\midrule
Observations&      196554&      196424&      196424&      196423\\
R-squared   &       0.447&       0.471&       0.473&       0.480\\
\midrule
Mean Dep. Var.&       0.606&       0.606&       0.606&       0.606\\
\bottomrule
\multicolumn{5}{l}{\footnotesize Standard errors in parentheses and clustered at the household level.}\\
\multicolumn{5}{l}{\footnotesize Dep. Var. Mean represents the dependent variable mean. Treatment is Benin Females after 2008}\\
\end{tabular}

\end{table}
\end{center}

\begin{center}
    \begin{table}
        \centering
\caption{Effect of free cesarean section on the probability of having more than 2 children \label{fertility4}}
\begin{tabular}{l*{4}{c}}
\toprule
            &\multicolumn{4}{c}{Having more than 2 children: 1 if yes, 0 otherwise}\\                     &\multicolumn{1}{c}{(1)}&\multicolumn{1}{c}{(2)}&\multicolumn{1}{c}{(3)}&\multicolumn{1}{c}{(4)}\\
\midrule
Treatment   &     -0.0157&     -0.0135&     -0.0181&     -0.0178\\
            &   (0.00447)&   (0.00445)&   (0.00443)&   (0.00443)\\
\addlinespace
Education Attainment&     -0.0685&     -0.0669&     -0.0556&     -0.0558\\
            &  (0.000531)&  (0.000533)&  (0.000602)&  (0.000607)\\
\addlinespace
Age of Female&      0.0337&      0.0343&      0.0345&      0.0346\\
            & (0.0000825)& (0.0000812)& (0.0000810)& (0.0000906)\\
\addlinespace
Age of Partner&            &    -0.00256&    -0.00258&    -0.00262\\
            &            & (0.0000574)& (0.0000573)& (0.0000589)\\
\addlinespace
Wealth Index&            &            &     -0.0257&     -0.0257\\
            &            &            &  (0.000654)&  (0.000655)\\
\addlinespace
Marital Status&            &            &            &    -0.00381\\
            &            &            &            &   (0.00114)\\
\addlinespace
Country Fixed Effect &   Yes   &  Yes      & Yes      &    Yes\\
\addlinespace
Time Fixed Effect.  &   Yes   &  Yes      & Yes      &    Yes\\
\addlinespace
\_cons      &      -0.370&      -0.277&      -0.214&      -0.212\\
            &   (0.00271)&   (0.00370)&   (0.00407)&   (0.00411)\\
\midrule
Observations&      196554&      196424&      196424&      196423\\
R-squared   &       0.469&       0.474&       0.478&       0.478\\
\midrule
Mean Dep. Var.&       0.487&       0.487&       0.487&       0.487\\
\bottomrule
\multicolumn{5}{l}{\footnotesize Standard errors in parentheses and clustered at the household level.}\\
\multicolumn{5}{l}{\footnotesize Dep. Var. Mean represents the dependent variable mean. Treatment is Benin Females after 2008}\\
\end{tabular}
    \end{table}
\end{center}

\begin{center}
    \begin{table}
        \centering
\caption{Marginal effects of free cesarean section \label{fertility5}}
\begin{tabular}{l*{3}{c}}
\toprule
            &\multicolumn{1}{c}{$1^{st}$ child}&\multicolumn{1}{c}{$2^{nd}$ child}&\multicolumn{1}{c}{$>$ 2 children}\\
\midrule
Treatment   &     0.0247&     -0.0164&     -0.0178\\
            &   (0.00367)&   (0.00418)&   (0.00443)\\
\addlinespace
Education Attainment&     -0.0507&     -0.0578&     -0.0558\\
            &  (0.000581)&  (0.000613)&  (0.000607)\\
\addlinespace
Age of Female&      0.0228&      0.0312&      0.0346\\
            & (0.0000882)& (0.0000930)& (0.0000906)\\
\addlinespace
Age of Partner&    -0.00622&    -0.00496&    -0.00262\\
            & (0.0000614)& (0.0000627)& (0.0000589)\\
\addlinespace
Wealth Index&     -0.0144&     -0.0191&     -0.0257\\
            &  (0.000580)&  (0.000636)&  (0.000655)\\
\addlinespace
Marital Status&       0.118&      0.0506&    -0.00381\\
            &   (0.00120)&   (0.00125)&   (0.00114)\\
\addlinespace
Country Fixed Effect &   Yes   &  Yes      & Yes      \\
\addlinespace
Time Fixed Effect.  &   Yes   &  Yes      & Yes      \\
\addlinespace
\_cons      &       0.365&      0.0338&      -0.212\\
            &   (0.00390)&   (0.00425)&   (0.00411)\\
\midrule
Observations&      196423&      196423&      196423\\
R-squared   &       0.468&       0.480&       0.478\\
\midrule
Mean Dep. Var.&       0.733&       0.606&       0.487\\
\bottomrule
\multicolumn{4}{l}{\footnotesize Standard errors in parentheses and clustered at the household level.}\\
\multicolumn{4}{l}{\footnotesize Dep. Var. Mean represents the dependent variable mean.}\\
\multicolumn{4}{l}{\footnotesize Treatment is Benin Females after 2008}\\
\end{tabular}

    \end{table}
\end{center}

\begin{center}
    \begin{table}
        \centering

    \end{table}
\end{center}

\end{document}